\pdftrailerid{}
\documentclass{article}

\usepackage{iclr2027/iclr2027_conference,times}
\usepackage[T1]{fontenc}
\usepackage[utf8]{inputenc}
\usepackage{amsmath,amssymb,mathtools,bm}
\DeclareMathSizes{9.2}{9.2}{6}{5}
\DeclareMathSizes{7.5}{7}{5.25}{5}
\DeclareMathSizes{6.5}{6}{5}{5}
\usepackage{graphicx}
\usepackage{booktabs}
\usepackage{array}
\usepackage{colortbl}
\usepackage{placeins}
\usepackage{needspace}
\usepackage{longtable}
\usepackage{microtype}
\usepackage{hyperref}
\usepackage{url}
\usepackage{fontawesome5}
\usepackage{xcolor}
\definecolor{capSequencePriors}{HTML}{9C7524}
\definecolor{capFunctionalPrediction}{HTML}{2463A6}
\definecolor{capBiologicalContext}{HTML}{19836E}
\definecolor{capGeneticInteractions}{HTML}{7956A5}
\definecolor{capGenomicDesign}{HTML}{BE4D35}
\definecolor{capMutationEffectsAndContext}{HTML}{19836E}
\definecolor{capEditExecution}{HTML}{526674}
\definecolor{capPersonalPrediction}{HTML}{AE4275}

\colorlet{capDesignDecisions}{capGenomicDesign}

\colorlet{capPrior}{capSequencePriors}

\colorlet{capSimulate}{capFunctionalPrediction}

\colorlet{capContext}{capBiologicalContext}

\colorlet{capCompose}{capGeneticInteractions}

\colorlet{capControl}{capGenomicDesign}

\colorlet{capResponse}{capMutationEffectsAndContext}

\colorlet{capExecution}{capEditExecution}

\colorlet{capPersonal}{capPersonalPrediction}

\newcommand{\caplabel}[1]{\csname caplabel#1\endcsname}

\DeclareRobustCommand{\capheading}[2]{\capname{#1}: #2}
\newcommand{\caplead}[2]{\textbf{#2}}
\newsavebox{\capPlotBox}
\newsavebox{\capLabelBox}
\newlength{\capPlotHeight}
\newlength{\capUpperRule}
\newlength{\capLowerRule}
\newcommand{\capfigure}[1]{%
  \begin{lrbox}{\capPlotBox}%
  \begin{minipage}[b]{\dimexpr\linewidth-16pt\relax}\centering}
\newcommand{\captable}[2][]{%
  \def\capTableHeight{#1}%
  \ifx\capTableHeight\empty
    \capfigure{#2}%
  \else
    \begin{lrbox}{\capPlotBox}%
    \begin{minipage}[b][#1][c]{\dimexpr\linewidth-16pt\relax}\centering
  \fi}
\newcommand{\caprender}[1]{%
  \end{minipage}\end{lrbox}%
  \sbox{\capLabelBox}{\rotatebox{90}{%
    \textcolor{cap#1}{\fontfamily{phv}\fontseries{m}\fontshape{sl}%
    \fontsize{7}{8}\selectfont \caplabel{#1}}}}%
  \setlength{\capPlotHeight}{\dimexpr\ht\capPlotBox+\dp\capPlotBox\relax}%
  \setlength{\capUpperRule}{15.07bp}%
  \setlength{\capLowerRule}{\dimexpr\capPlotHeight-\capUpperRule-\ht\capLabelBox-\dp\capLabelBox-11pt\relax}%
  \par\noindent\hbox to\linewidth{\hfil%
    \raisebox{\dp\capPlotBox}{\usebox{\capPlotBox}}\hspace{6pt}%
    \vbox{\offinterlineskip%
      \hbox to8pt{\hfil\textcolor{cap#1!80!white}{\rule{.45pt}{\capUpperRule}}\hfil}%
      \vskip5.5pt%
      \hbox to8pt{\hfil\usebox{\capLabelBox}\hfil}%
      \vskip5.5pt%
      \hbox to8pt{\hfil\textcolor{cap#1!80!white}{\rule{.45pt}{\capLowerRule}}\hfil}}%
    \hfil}\par}
\newcommand{\capcaption}[2]{\caprender{#1}%
  \setlength{\abovecaptionskip}{4pt}\caption{#2}}

\DeclareRobustCommand{\caphighlight}[1]{%
  \begingroup\setlength{\fboxsep}{1.4pt}%
  \colorbox{cap#1!17!white}{\textcolor{black}{\caplabel{#1}:}}\endgroup}
\graphicspath{{figures/}}
\setkeys{Gin}{keepaspectratio}
\newcommand{\G}{\mathcal{G}}

\DeclareMathOperator*{\argmin}{arg\,min}

\title{BEND 2.0: Benchmarking the Capabilities\\ Needed to Design DNA}
\input{author-affiliations}
\hypersetup{
pdftitle={BEND 2.0: Benchmarking the Capabilities Needed to Design DNA}, pdfauthor={Andreas Bjerregaard; Rachael DeVries; Dionysia Danai Brilli; Panagiotis Antoniadis; Yan Li; Anders Krogh; Wouter Boomsma; Ole Winther; Frederikke Isa Marin}
}
\input{preprint-style}

\iclrpreprintcopy

\begin{document}
\maketitle

\begin{abstract}
Genomic foundation models are increasingly used to propose DNA sequences and edits. But successful design requires capabilities that current benchmarks do not measure. Here we introduce BEND 2.0 (\underline{BEN}chmark for \underline{D}NA language models), which tests how well models predict the effects of DNA changes and whether these predictions can guide sequence design. Across eight tasks built on five experimental datasets, we find that model choice and how representations are accessed both affect performance. Linear readouts fitted to frozen representations can match or exceed models' native outputs, and functional post-training substantially improves the models'  capabilities. Linear readouts transfer well across experimental assays and genes, suggesting that models capture functional patterns beyond the data used to fit individual predictors. At the same time, strong prediction of average effects can conceal weaker prediction of context-specific effects, and accurate prediction of combined mutation effects does not imply accurate prediction of their interactions. In design tasks, model predictions can reduce the number of candidates needed to find a desired response, but still leave substantial room for better design choices, particularly for context-specific and multi-cell objectives. Our benchmark reveals that genomic models now display signs of general utility but have yet to saturate on performance in even simple tasks.

\end{abstract}

\definecolor{resourceLink}{HTML}{8C493F}
\definecolor{resourceIcon}{HTML}{77736F}
\definecolor{resourceLabel}{HTML}{34302E}

\newcommand{\resourceicon}[2]{%
  \makebox[14pt][c]{\raisebox{\dimexpr3.1pt-.5\height\relax}[8pt][2pt]{%
    \resizebox{!}{#1}{#2}}}}

\newcommand{\resourceentry}[5]{%
  \textcolor{resourceIcon}{#1} &
  \textcolor{resourceLabel}{\fontsize{8.8}{11}\selectfont\textls[35]{\textsc{#2}}} &
  \href{#3}{\textcolor{resourceLink}{\nolinkurl{#4#5}}}\\[2.5pt]}

\par\vspace{-7pt}
\begingroup
\centering
\hypersetup{hidelinks}
\urlstyle{same}
\fontsize{9.2}{11}\selectfont
\setlength{\tabcolsep}{0pt}
\renewcommand{\arraystretch}{1}
\begin{tabular}{@{}l@{\hspace{6pt}}l@{\hspace{10pt}}l@{}}
\resourceentry{\resourceicon{7.8pt}{\faListOl}}
  {Leaderboard}{https://machinelearninglifescience.github.io/BEND2/}{machinelearninglifescience.github.io/}{BEND2}
\resourceentry{\resourceicon{8.6pt}{\faGithub}}
  {Code}{https://github.com/machinelearninglifescience/BEND2}{github.com/machinelearninglifescience/}{BEND2}
\resourceentry{\resourceicon{9.4pt}{\includegraphics{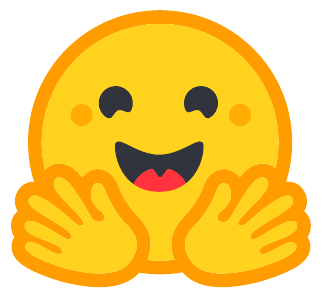}}}
  {Data}{https://huggingface.co/datasets/yhsure/BEND2}{huggingface.co/datasets/yhsure/}{BEND2}
\end{tabular}\par
\endgroup
\vspace{-7pt}

\section{Introduction}
\vspace{2.8pt}
A biologist seeking to change the activity of a regulatory sequence may have hundreds of mutations to choose from. The same mutation can have different effects in different cell types or even on a slightly different genetic background. To help choose what to test, models must distinguish small changes most likely to produce a fine-grained desired response.

Genomic models use different approaches to learn the complex and multifaceted patterns encoded in DNA, drawing on both sequence diversity and measurements of biological function. Models such as Evo2 and DNABERT-2 learn from natural sequence variation \citep{brixi2026evo2,zhou2024dnabert2}, while GPN-Star uses evolutionary relationships to predict functional constraint \citep{ye2025predicting}. Other models incorporate functional outcomes rather than relying on natural sequence variation and homology. AlphaGenome, for example, predicts expression, accessibility and splicing from functional genomic data \citep{avsec2026alphagenome}, while NTv3 combines sequence learning with such functional supervision \citep{boshar2025ntv3}. Foundation models and task-specific models have guided the design of sequences with cell type-targeted activity \citep{gosai2024design} and, more recently, viable bacteriophage genomes \citep{king2026generative}. However, the central question remains: which features of sequence and function have these models learned well enough to guide experiments?

Looking only at overall predictive accuracy leaves that question unanswered. A sequence-based score may identify a mutation at an evolutionarily constrained position without predicting its effect on expression in a particular cell type. Accurate predictions for individual mutations do not establish whether a model captures interactions between mutations or can predict their effects on a different genetic background. Similarly, distinguishing highly active genes from weakly active genes does not indicate which edit will produce a desired change in expression. Each of these cases calls for more precise testing.

Building on BEND~\citep{marin2024bend}, we introduce BEND 2.0 to test whether genomic models can guide DNA design. The benchmark is organized around five key capabilities: %
\textcolor{capSequencePriors}{Prior},
\textcolor{capFunctionalPrediction}{Function},
\textcolor{capBiologicalContext}{Context},
\textcolor{capGeneticInteractions}{Interaction}, and
\textcolor{capGenomicDesign}{Design}.
Together, our tasks measure what genomic models can predict and whether they can help select DNA sequences that produce a desired response. Table~\ref{tab:bend2-reference} summarizes our selected evaluations.

\paragraph{Related works.}

The original BEND benchmark \citep{marin2024bend}, GenBench \citep{liu2024genbench}, and OmniGenBench \citep{yang2025omnigenbench} provide systematic evaluations of genomic representations across downstream tasks. In parallel, model-guided regulatory sequence design has demonstrated cell type-targeted activity experimentally \citep{gosai2024design,yin2025enhancer,castillohair2025atlas}, and NucleoBench \citep{shor2025nucleobench} evaluates algorithms for searching sequence space using predictive models as scoring functions. BEND 2.0 connects these directions by testing the predictive distinctions that highly matter for design: resolving context-specific effects, predicting interactions between mutations, and distinguishing candidates with different functional outcomes. %

\begin{table}[!ht]
\vspace{6pt}
\centering\fontfamily{phv}\selectfont\fontsize{7.5}{8.5}\selectfont
\hypersetup{hidelinks}
\setlength{\tabcolsep}{1.5pt}\renewcommand{\arraystretch}{1.065}
\definecolor{benchInk}{HTML}{26323D}
\definecolor{benchRail}{HTML}{FFFFFF}\definecolor{benchRailHead}{HTML}{FFFFFF}
\definecolor{benchPrior}{HTML}{A18338}
\definecolor{benchSimulate}{HTML}{477FA5}
\definecolor{benchContext}{HTML}{438879}
\definecolor{benchCompose}{HTML}{8266A2}
\definecolor{benchControl}{HTML}{B56E58}
\color{benchInk}
\newcommand{\benchhead}[2]{\textcolor{bench#1!85!black}{\mdseries\fontsize{6.8}{8.5}\selectfont #2}}
\newcommand{\benchmetric}[1]{{\mdseries\fontsize{7.2}{8.7}\selectfont\shortstack{#1}}}
\newcommand{\benchentry}[1]{\raisebox{-0.75pt}[0pt][0pt]{#1}}
\begin{tabular}{@{}>{\columncolor{benchRail}[0pt][0pt]\raggedright\arraybackslash}p{\dimexpr.205\linewidth-1.5pt\relax}>{\columncolor{benchPrior!3}[0pt][0pt]\raggedleft\arraybackslash}p{\dimexpr.091\linewidth-3pt\relax}>{\columncolor{benchSimulate!3}[0pt][0pt]\raggedleft\arraybackslash}p{\dimexpr.088\linewidth-3pt\relax}>{\columncolor{benchContext!3}[0pt][1.5pt]\raggedleft\arraybackslash}p{\dimexpr.105\linewidth-3pt\relax}>{\columncolor{benchContext!3}[1.5pt][0pt]\raggedleft\arraybackslash}p{\dimexpr.085\linewidth-3pt\relax}>{\columncolor{benchCompose!3}[0pt][0pt]\raggedleft\arraybackslash}p{\dimexpr.11\linewidth-3pt\relax}>{\columncolor{benchControl!3}[0pt][1.5pt]\raggedleft\arraybackslash}p{\dimexpr.10\linewidth-3pt\relax}>{\columncolor{benchControl!3}[1.5pt][1.5pt]\raggedleft\arraybackslash}p{\dimexpr.111\linewidth-3pt\relax}>{\columncolor{benchControl!3}[1.5pt][0pt]\raggedleft\arraybackslash}p{\dimexpr.105\linewidth-3pt\relax}@{}}
\multicolumn{1}{@{}>{\columncolor{benchRailHead}[0pt][0pt]}p{\dimexpr.205\linewidth-1.5pt\relax}}{\rule{0pt}{13pt}\par\vspace{4pt}\raisebox{-16pt}[0pt][0pt]{{\fontsize{7.5}{8.5}\selectfont\bfseries Model}}} & \multicolumn{1}{>{\columncolor{benchPrior!6}[0pt][0pt]}p{\dimexpr.091\linewidth-3pt\relax}}{\centering\arraybackslash\rule{0pt}{13pt}\benchhead{Prior}{\hyperref[fig:prior]{Fig.\,\ref*{fig:prior}A}}\par\vspace{4pt}\benchmetric{Prior\\loss\\{\color{benchInk!80}\fontsize{6.5}{8}\selectfont bits $\downarrow$}}\rule[-5pt]{0pt}{5pt}} & \multicolumn{1}{>{\columncolor{benchSimulate!6}[0pt][0pt]}p{\dimexpr.088\linewidth-3pt\relax}}{\centering\arraybackslash\rule{0pt}{13pt}\benchhead{Simulate}{\hyperref[fig:simulate]{Fig.\,\ref*{fig:simulate}A}}\par\vspace{4pt}\benchmetric{Splicing\\effects\\{\color{benchInk!80}\fontsize{6.5}{8}\selectfont $\rho\uparrow$}}\rule[-5pt]{0pt}{5pt}} & \multicolumn{1}{>{\columncolor{benchContext!6}[0pt][1.5pt]}p{\dimexpr.105\linewidth-3pt\relax}}{\centering\arraybackslash\rule{0pt}{13pt}\benchhead{Context}{\hyperref[fig:context]{Fig.\,\ref*{fig:context}B}}\par\vspace{4pt}\benchmetric{Cell\\contrast\\{\color{benchInk!80}\fontsize{6.5}{8}\selectfont $\rho\uparrow$}}\rule[-5pt]{0pt}{5pt}} & \multicolumn{1}{>{\columncolor{benchContext!6}[1.5pt][0pt]}p{\dimexpr.085\linewidth-3pt\relax}}{\centering\arraybackslash\rule{0pt}{13pt}\benchhead{Context}{\hyperref[fig:context-tissues]{Fig.\,\ref*{fig:context-tissues}B}}\par\vspace{4pt}\benchmetric{Tissue\\ID\\{\color{benchInk!80}\fontsize{6.5}{8}\selectfont \% $\uparrow$}}\rule[-5pt]{0pt}{5pt}} & \multicolumn{1}{>{\columncolor{benchCompose!6}[0pt][0pt]}p{\dimexpr.11\linewidth-3pt\relax}}{\centering\arraybackslash\rule{0pt}{13pt}\benchhead{Compose}{\hyperref[fig:compose]{Fig.\,\ref*{fig:compose}A}}\par\vspace{4pt}\benchmetric{Background\\interaction\\{\color{benchInk!80}\fontsize{6.5}{8}\selectfont $\rho\uparrow$}}\rule[-5pt]{0pt}{5pt}} & \multicolumn{1}{>{\columncolor{benchControl!6}[0pt][1.5pt]}p{\dimexpr.10\linewidth-3pt\relax}}{\centering\arraybackslash\rule{0pt}{13pt}\benchhead{Control}{\hyperref[fig:control]{Fig.\,\ref*{fig:control}A}}\par\vspace{4pt}\benchmetric{Single-edit\\rank\\{\color{benchInk!80}\fontsize{6.5}{8}\selectfont percentile $\downarrow$}}\rule[-5pt]{0pt}{5pt}} & \multicolumn{1}{>{\columncolor{benchControl!6}[1.5pt][1.5pt]}p{\dimexpr.111\linewidth-3pt\relax}}{\centering\arraybackslash\rule{0pt}{13pt}\benchhead{Control}{\hyperref[fig:control]{Fig.\,\ref*{fig:control}C}}\par\vspace{4pt}\benchmetric{Cell\\selectivity\\{\color{benchInk!80}\fontsize{6.5}{8}\selectfont $U\uparrow$}}\rule[-5pt]{0pt}{5pt}} & \multicolumn{1}{>{\columncolor{benchControl!6}[1.5pt][0pt]}p{\dimexpr.105\linewidth-3pt\relax}}{\centering\arraybackslash\rule{0pt}{13pt}\benchhead{Control}{\hyperref[fig:frontier-programming]{Fig.\,\ref*{fig:frontier-programming}A}}\par\vspace{4pt}\benchmetric{Multi-cell\\gain\\{\color{benchInk!80}\fontsize{6.5}{8}\selectfont log$_2$ $\uparrow$}}\rule[-5pt]{0pt}{5pt}}\\[3pt]
\benchentry{Base composition} & \cellcolor{benchPrior!5.00}\benchentry{\makebox[\linewidth][c]{1.974}} & \cellcolor{benchSimulate!5.00}\benchentry{\makebox[\linewidth][c]{\textcolor{benchInk!27}{---}}} & \cellcolor{benchContext!5.00}\benchentry{\makebox[\linewidth][c]{\textcolor{benchInk!27}{---}}} & \cellcolor{benchContext!5.00}\benchentry{\makebox[\linewidth][c]{\textcolor{benchInk!27}{---}}} & \cellcolor{benchCompose!5.00}\benchentry{\makebox[\linewidth][c]{\textcolor{benchInk!27}{---}}} & \cellcolor{benchControl!5.00}\benchentry{\makebox[\linewidth][c]{\textcolor{benchInk!27}{---}}} & \cellcolor{benchControl!5.00}\benchentry{\makebox[\linewidth][c]{\textcolor{benchInk!27}{---}}} & \cellcolor{benchControl!5.00}\benchentry{\makebox[\linewidth][c]{\textcolor{benchInk!27}{---}}}\\
\benchentry{Evo2 20B} & \cellcolor{benchPrior!31.47}\benchentry{\makebox[\linewidth][c]{1.388}} & \cellcolor{benchSimulate!6.86}\benchentry{\makebox[\linewidth][c]{0.367}} & \cellcolor{benchContext!5.39}\benchentry{\makebox[\linewidth][c]{0.041}} & \cellcolor{benchContext!6.98}\benchentry{\makebox[\linewidth][c]{21.6}} & \cellcolor{benchCompose!6.92}\benchentry{\makebox[\linewidth][c]{0.050}} & \cellcolor{benchControl!7.27}\benchentry{\makebox[\linewidth][c]{43.5}} & \cellcolor{benchControl!5.00}\benchentry{\makebox[\linewidth][c]{\llap{-}0.350}} & \cellcolor{benchControl!7.63}\benchentry{\makebox[\linewidth][c]{0.275}}\\
\benchentry{CENO 1B} & \cellcolor{benchPrior!7.99}\benchentry{\makebox[\linewidth][c]{1.549}} & \cellcolor{benchSimulate!5.56}\benchentry{\makebox[\linewidth][c]{0.295}} & \cellcolor{benchContext!5.16}\benchentry{\makebox[\linewidth][c]{0.036}} & \cellcolor{benchContext!5.70}\benchentry{\makebox[\linewidth][c]{\phantom{0}9.8}} & \cellcolor{benchCompose!5.16}\benchentry{\makebox[\linewidth][c]{0.010}} & \cellcolor{benchControl!5.03}\benchentry{\makebox[\linewidth][c]{48.9}} & \cellcolor{benchControl!5.03}\benchentry{\makebox[\linewidth][c]{\llap{-}0.318}} & \cellcolor{benchControl!5.00}\benchentry{\makebox[\linewidth][c]{\llap{-}0.581}}\\
\benchentry{GENERATORv2} & \cellcolor{benchPrior!5.00}\benchentry{\makebox[\linewidth][c]{1.711}} & \cellcolor{benchSimulate!7.70}\benchentry{\makebox[\linewidth][c]{0.388}} & \cellcolor{benchContext!5.78}\benchentry{\makebox[\linewidth][c]{0.066}} & \cellcolor{benchContext!6.45}\benchentry{\makebox[\linewidth][c]{17.6}} & \cellcolor{benchCompose!6.29}\benchentry{\makebox[\linewidth][c]{0.046}} & \cellcolor{benchControl!6.57}\benchentry{\makebox[\linewidth][c]{44.7}} & \cellcolor{benchControl!5.32}\benchentry{\makebox[\linewidth][c]{\llap{-}0.256}} & \cellcolor{benchControl!7.03}\benchentry{\makebox[\linewidth][c]{0.234}}\\
\benchentry{DNABERT-2} & \cellcolor{benchPrior!5.00}\benchentry{\makebox[\linewidth][c]{\textcolor{benchInk!27}{---}}} & \cellcolor{benchSimulate!5.16}\benchentry{\makebox[\linewidth][c]{0.198}} & \cellcolor{benchContext!6.28}\benchentry{\makebox[\linewidth][c]{0.076}} & \cellcolor{benchContext!5.06}\benchentry{\makebox[\linewidth][c]{\phantom{0}5.9}} & \cellcolor{benchCompose!5.39}\benchentry{\makebox[\linewidth][c]{0.020}} & \cellcolor{benchControl!5.13}\benchentry{\makebox[\linewidth][c]{48.6}} & \cellcolor{benchControl!5.13}\benchentry{\makebox[\linewidth][c]{\llap{-}0.302}} & \cellcolor{benchControl!6.79}\benchentry{\makebox[\linewidth][c]{0.233}}\\
\addlinespace[2.2pt]
\benchentry{NTv3 100M pre} & \cellcolor{benchPrior!11.06}\benchentry{\makebox[\linewidth][c]{1.537}} & \cellcolor{benchSimulate!6.21}\benchentry{\makebox[\linewidth][c]{0.349}} & \cellcolor{benchContext!6.28}\benchentry{\makebox[\linewidth][c]{0.076}} & \cellcolor{benchContext!5.28}\benchentry{\makebox[\linewidth][c]{\phantom{0}7.8}} & \cellcolor{benchCompose!5.16}\benchentry{\makebox[\linewidth][c]{0.010}} & \cellcolor{benchControl!5.32}\benchentry{\makebox[\linewidth][c]{47.3}} & \cellcolor{benchControl!6.02}\benchentry{\makebox[\linewidth][c]{\llap{-}0.207}} & \cellcolor{benchControl!8.54}\benchentry{\makebox[\linewidth][c]{0.339}}\\
\benchentry{NTv3 100M post} & \cellcolor{benchPrior!15.15}\benchentry{\makebox[\linewidth][c]{1.492}} & \cellcolor{benchSimulate!12.53}\benchentry{\makebox[\linewidth][c]{0.502}} & \cellcolor{benchContext!11.81}\benchentry{\makebox[\linewidth][c]{0.244}} & \cellcolor{benchContext!16.34}\benchentry{\makebox[\linewidth][c]{51.0}} & \cellcolor{benchCompose!9.05}\benchentry{\makebox[\linewidth][c]{0.139}} & \cellcolor{benchControl!14.14}\benchentry{\makebox[\linewidth][c]{30.3}} & \cellcolor{benchControl!11.38}\benchentry{\makebox[\linewidth][c]{0.211}} & \cellcolor{benchControl!6.18}\benchentry{\makebox[\linewidth][c]{0.144}}\\
\benchentry{\hspace{6pt}\textcolor{black!65}{Native outputs}} & \cellcolor{benchPrior!5.00}\benchentry{\makebox[\linewidth][c]{\textcolor{benchInk!27}{---}}} & \cellcolor{benchSimulate!5.00}\benchentry{\makebox[\linewidth][c]{0.168}} & \cellcolor{benchContext!8.06}\benchentry{\makebox[\linewidth][c]{0.203}} & \cellcolor{benchContext!7.76}\benchentry{\makebox[\linewidth][c]{27.5}} & \cellcolor{benchCompose!10.40}\benchentry{\makebox[\linewidth][c]{0.152}} & \cellcolor{benchControl!11.22}\benchentry{\makebox[\linewidth][c]{32.0}} & \cellcolor{benchControl!10.35}\benchentry{\makebox[\linewidth][c]{0.200}} & \cellcolor{benchControl!9.13}\benchentry{\makebox[\linewidth][c]{0.344\rlap{$^{\ddagger}$}}}\\
\addlinespace[1.8pt]
\benchentry{NTv3 650M pre} & \cellcolor{benchPrior!22.22}\benchentry{\makebox[\linewidth][c]{1.447}} & \cellcolor{benchSimulate!5.03}\benchentry{\makebox[\linewidth][c]{0.187}} & \cellcolor{benchContext!5.00}\benchentry{\makebox[\linewidth][c]{0.010}} & \cellcolor{benchContext!5.28}\benchentry{\makebox[\linewidth][c]{\phantom{0}7.8}} & \cellcolor{benchCompose!5.03}\benchentry{\makebox[\linewidth][c]{0.009}} & \cellcolor{benchControl!5.00}\benchentry{\makebox[\linewidth][c]{50.4}} & \cellcolor{benchControl!7.47}\benchentry{\makebox[\linewidth][c]{\llap{-}0.061}} & \cellcolor{benchControl!10.77}\benchentry{\makebox[\linewidth][c]{0.368}}\\
\benchentry{NTv3 650M post} & \cellcolor{benchPrior!38.00}\benchentry{\makebox[\linewidth][c]{1.375}} & \cellcolor{benchSimulate!17.14}\benchentry{\makebox[\linewidth][c]{0.564}} & \cellcolor{benchContext!14.85}\benchentry{\makebox[\linewidth][c]{0.256}} & \cellcolor{benchContext!22.97}\benchentry{\makebox[\linewidth][c]{56.9}} & \cellcolor{benchCompose!20.73}\benchentry{\makebox[\linewidth][c]{0.187}} & \cellcolor{benchControl!17.12}\benchentry{\makebox[\linewidth][c]{30.0}} & \cellcolor{benchControl!16.35}\benchentry{\makebox[\linewidth][c]{0.287}} & \cellcolor{benchControl!17.68}\benchentry{\makebox[\linewidth][c]{0.568}}\\
\benchentry{\hspace{6pt}\textcolor{black!65}{Native outputs}} & \cellcolor{benchPrior!5.00}\benchentry{\makebox[\linewidth][c]{\textcolor{benchInk!27}{---}}} & \cellcolor{benchSimulate!9.57}\benchentry{\makebox[\linewidth][c]{0.461}} & \cellcolor{benchContext!10.57}\benchentry{\makebox[\linewidth][c]{0.239}} & \cellcolor{benchContext!12.38}\benchentry{\makebox[\linewidth][c]{45.1}} & \cellcolor{benchCompose!12.20}\benchentry{\makebox[\linewidth][c]{0.170}} & \cellcolor{benchControl!22.65}\benchentry{\makebox[\linewidth][c]{29.3}} & \cellcolor{benchControl!12.98}\benchentry{\makebox[\linewidth][c]{0.250}} & \cellcolor{benchControl!13.84}\benchentry{\makebox[\linewidth][c]{0.482\rlap{$^{\ddagger}$}}}\\
\addlinespace[2.2pt]
\benchentry{AlphaGenome} & \cellcolor{benchPrior!5.00}\benchentry{\makebox[\linewidth][c]{\textcolor{benchInk!27}{---}}} & \cellcolor{benchSimulate!31.23}\benchentry{\makebox[\linewidth][c]{0.650}} & \cellcolor{benchContext!38.00}\benchentry{\makebox[\linewidth][c]{0.402}} & \cellcolor{benchContext!30.93}\benchentry{\makebox[\linewidth][c]{64.7}} & \cellcolor{benchCompose!38.00}\benchentry{\makebox[\linewidth][c]{0.269}} & \cellcolor{benchControl!31.06}\benchentry{\makebox[\linewidth][c]{25.7}} & \cellcolor{benchControl!38.00}\benchentry{\makebox[\linewidth][c]{0.515}} & \cellcolor{benchControl!38.00}\benchentry{\makebox[\linewidth][c]{0.833}}\\
\benchentry{\hspace{6pt}\textcolor{black!65}{Native outputs}} & \cellcolor{benchPrior!5.00}\benchentry{\makebox[\linewidth][c]{\textcolor{benchInk!27}{---}}} & \cellcolor{benchSimulate!24.51}\benchentry{\makebox[\linewidth][c]{0.621}} & \cellcolor{benchContext!32.27}\benchentry{\makebox[\linewidth][c]{0.398}} & \cellcolor{benchContext!38.00}\benchentry{\makebox[\linewidth][c]{68.6}} & \cellcolor{benchCompose!31.61}\benchentry{\makebox[\linewidth][c]{0.260}} & \cellcolor{benchControl!38.00}\benchentry{\makebox[\linewidth][c]{24.3}} & \cellcolor{benchControl!28.69}\benchentry{\makebox[\linewidth][c]{0.384}} & \cellcolor{benchControl!30.15}\benchentry{\makebox[\linewidth][c]{0.707\rlap{$^{\ddagger}$}}}\\
\benchentry{Enformer} & \cellcolor{benchPrior!5.00}\benchentry{\makebox[\linewidth][c]{\textcolor{benchInk!27}{---}}} & \cellcolor{benchSimulate!5.00}\benchentry{\makebox[\linewidth][c]{\textcolor{benchInk!27}{---}}} & \cellcolor{benchContext!20.69}\benchentry{\makebox[\linewidth][c]{0.282}} & \cellcolor{benchContext!9.17}\benchentry{\makebox[\linewidth][c]{31.4}} & \cellcolor{benchCompose!15.44}\benchentry{\makebox[\linewidth][c]{0.181}} & \cellcolor{benchControl!12.39}\benchentry{\makebox[\linewidth][c]{31.5}} & \cellcolor{benchControl!21.41}\benchentry{\makebox[\linewidth][c]{0.290}} & \cellcolor{benchControl!24.76}\benchentry{\makebox[\linewidth][c]{0.700\rlap{$^{\ddagger}$}}}\\
\addlinespace[2.2pt]
\benchentry{$k$-mer ridge} & \cellcolor{benchPrior!5.00}\benchentry{\makebox[\linewidth][c]{\textcolor{benchInk!27}{---}}} & \cellcolor{benchSimulate!5.65}\benchentry{\makebox[\linewidth][c]{0.318}} & \cellcolor{benchContext!9.00}\benchentry{\makebox[\linewidth][c]{0.211}} & \cellcolor{benchContext!5.28}\benchentry{\makebox[\linewidth][c]{\phantom{0}7.8}} & \cellcolor{benchCompose!5.76}\benchentry{\makebox[\linewidth][c]{0.032}} & \cellcolor{benchControl!6.07}\benchentry{\makebox[\linewidth][c]{45.0}} & \cellcolor{benchControl!6.73}\benchentry{\makebox[\linewidth][c]{\llap{-}0.097}} & \cellcolor{benchControl!11.91}\benchentry{\makebox[\linewidth][c]{0.394}}\\
\benchentry{Task-trained ResNet} & \cellcolor{benchPrior!5.00}\benchentry{\makebox[\linewidth][c]{\textcolor{benchInk!27}{---}}} & \cellcolor{benchSimulate!11.09}\benchentry{\makebox[\linewidth][c]{0.488}} & \cellcolor{benchContext!5.03}\benchentry{\makebox[\linewidth][c]{0.020}} & \cellcolor{benchContext!5.00}\benchentry{\makebox[\linewidth][c]{\phantom{0}2.0}} & \cellcolor{benchCompose!5.00}\benchentry{\makebox[\linewidth][c]{\llap{-}0.001}} & \cellcolor{benchControl!5.61}\benchentry{\makebox[\linewidth][c]{47.0}} & \cellcolor{benchControl!5.63}\benchentry{\makebox[\linewidth][c]{\llap{-}0.215}} & \cellcolor{benchControl!9.82}\benchentry{\makebox[\linewidth][c]{0.347}}\\
\benchentry{SpliceAI} & \cellcolor{benchPrior!5.00}\benchentry{\makebox[\linewidth][c]{\textcolor{benchInk!27}{---}}} & \cellcolor{benchSimulate!38.00}\benchentry{\makebox[\linewidth][c]{0.674}} & \cellcolor{benchContext!5.00}\benchentry{\makebox[\linewidth][c]{\textcolor{benchInk!27}{---}}} & \cellcolor{benchContext!5.00}\benchentry{\makebox[\linewidth][c]{\textcolor{benchInk!27}{---}}} & \cellcolor{benchCompose!5.00}\benchentry{\makebox[\linewidth][c]{\textcolor{benchInk!27}{---}}} & \cellcolor{benchControl!5.00}\benchentry{\makebox[\linewidth][c]{\textcolor{benchInk!27}{---}}} & \cellcolor{benchControl!5.00}\benchentry{\makebox[\linewidth][c]{\textcolor{benchInk!27}{---}}} & \cellcolor{benchControl!5.00}\benchentry{\makebox[\linewidth][c]{\textcolor{benchInk!27}{---}}}\\
\addlinespace[3pt]
\rowcolor{black!3}\cellcolor{white}\benchentry{\textcolor{black!65}{\fontsize{7.2}{8.5}\selectfont Experimental reference\rlap{$^{\dagger}$}}} & \benchentry{\makebox[\linewidth][c]{\textcolor{benchInk!27}{---}}} & \benchentry{\makebox[\linewidth][c]{0.809}} & \benchentry{\makebox[\linewidth][c]{0.594}} & \benchentry{\makebox[\linewidth][c]{99.0}} & \benchentry{\makebox[\linewidth][c]{0.421}} & \benchentry{\makebox[\linewidth][c]{11.8}} & \benchentry{\makebox[\linewidth][c]{0.962}} & \benchentry{\makebox[\linewidth][c]{1.602}}\\
\end{tabular}
\color{black}\rmfamily\setlength{\abovecaptionskip}{4pt}
\caption{\textbf{The BEND 2.0 benchmark.} Metrics across categories collected in one overview table. The referenced figures and their corresponding sections provide further exploration. Deeper shading indicates better performance within each column.
Results use embedding-fitted linear heads, except for prior loss (native sequence logits) and Enformer/indented rows (native functional outputs).
$\ddagger$: linear head fitted to native outputs.
$\dagger$: Experimental references use experimental subsets but can be naive, e.g., \textit{Background interaction} uses a simple transfer from HepG2 to K562 (see Appendix~\hyperref[app:experimental-references]{\ref*{app:predictions-readouts}}).
}
\label{tab:bend2-reference}
\vspace{-6pt}
\end{table}

\FloatBarrier
\vspace{6pt}

\section{A framework for evaluating genomic models}
\label{sec:questions}

BEND 2.0 is centered around five complementary capabilities that connect sequence and functional understanding to genomic design.

\begingroup
\setlength{\abovedisplayskip}{5pt plus 1pt minus 2pt}
\setlength{\belowdisplayskip}{5pt plus 1pt minus 2pt}
\setlength{\abovedisplayshortskip}{0pt plus 1pt}
\setlength{\belowdisplayshortskip}{3pt plus 1pt minus 1pt}
\paragraph{\caphighlight{SequencePriors} How probable is the DNA?}
We measure how well a model predicts a masked base $x_i$ from surrounding DNA:
\begin{equation}
    p_\theta\!\left(x_i \mid \text{available DNA context}\right)
    \label{eq:sequence-prior}
\end{equation}
This is computed via the negative log-probability, $-\log_2 p_\theta(x_i\mid\text{available DNA context})$, over all queried positions. We report bits per base where lower values indicate better predictions.

\Needspace{7\baselineskip}
\paragraph{\caphighlight{FunctionalPrediction} Can sequence predict a functional outcome?}
For a reference sequence $x$ and an edit $\delta$, the mutation effect $\Delta y$ is the change in an experimental measurement between $x$ and $x+\delta$. A predictor $f_\theta$ estimates this effect:
\begin{equation}
    (x,\;x+\delta)
    \;\xrightarrow{\;f_\theta\;}\;
    \widehat{\Delta y}
    \label{eq:functional-prediction}
\end{equation}
We test mutation-effect predictions on held-out genes, then evaluate whether the fitted outputs transfer to another assay. %

\paragraph{\caphighlight{BiologicalContext} How does genomic function depend on the context?}
For a DNA sequence with an experimental measurement $y_c$ measured in several cells or tissues $c$, we separate its mean across the evaluated contexts $\bar y$ from the context-specific deviation $\kappa_c$:
\begin{equation}
    y_c
    =
    \underbrace{\bar y}_{\text{mean across contexts}}
    +
    \underbrace{\kappa_c}_{\text{context-specific deviation}}
    \label{eq:context}
\end{equation}
We test mutation-effect differences between cells, expression differences between tissues, and whether predicted profiles across held-out genes identify the correct tissue.

\paragraph{\caphighlight{GeneticInteractions} Does one mutation alter another's effect?}
Let $\Delta y_1$ and $\Delta y_2$ denote the effects of two individual mutations, and $\Delta y_{1+2}$ their combined effect, all relative to a reference sequence in the same context. The interaction $\mathcal I_{1,2}$ is the non-additive part of the signal:
\begin{equation}
    \Delta y_{1+2}
    =
    \underbrace{\Delta y_1+\Delta y_2}_{\text{additive expectation}}
    +
    \underbrace{\mathcal I_{1,2}}_{\text{interaction}}
    \label{eq:compose}
\end{equation}
We hence score interaction separately since an accurate combined effect can mask poor predictions of interaction.

\paragraph{\caphighlight{GenomicDesign} Which candidate achieves the desired response?}
Given a target response $\mathbf y_{\mathrm{target}}$, the model chooses a candidate $a^\star$ among permitted edits or sequences $\mathcal C$  whose predicted response $\widehat{\mathbf y}(a)$ best matches the target $\mathbf y_{\mathrm{target}}$:
\begin{equation}
    a^\star
    =
    \operatorname*{arg\,min}_{a\in\mathcal C}
    \ell\!\left(
        \widehat{\mathbf y}(a),\,
        \mathbf y_{\mathrm{target}}
    \right)
    \label{eq:control}
\end{equation}
The loss $\ell$ measures mismatch with the target, using the responses relevant to each task: mutation effects or activity across cells.

\endgroup

\paragraph{Predictions and generalization.}
Most predictions use linear heads fitted to frozen model sequence embeddings. For mutation effects, the linear head weights $w$ are fitted to the difference between variant and reference embeddings ($\phi_\theta(x)$) with $\widehat{\Delta y} =w^\top[\phi_\theta(x+\delta)-\phi_\theta(x)]$. These predictions and native outputs are reported separately. Detailed descriptions of the output types can be found in Appendix \ref{app:predictions-readouts}. We compute correlations within each gene or regulatory sequence, then average across the $G$ groups using $\rho_{\mathrm{macro}}=G^{-1}\sum_{g=1}^{G}\rho_g$. This avoids the score being dominated by, e.g., one large gene. For evaluation splits, we group related examples by gene, regulatory element, locus or sequence family. In addition to genomic foundation models, our comparisons include simple predictors, specialist models and experimental references where applicable.

\section{\capheading{SequencePriors}{measuring what models know about DNA}} \label{sec:sequence-priors}

\begin{figure}[!tb]
 \setlength{\parskip}{0pt}
 \centering\capfigure{SequencePriors}
 \includegraphics[trim=0bp 2bp 0bp 4bp,clip,width=.97\linewidth]{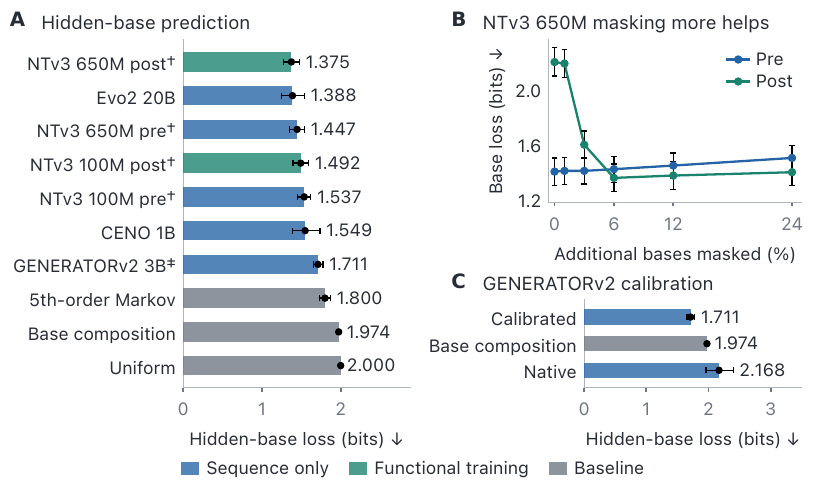}
 \capcaption{SequencePriors}{\textbf{Sequence-prior performance depends on the representation extraction.}
 A: masked loss on 512 windows. NTv3 uses fixed 12\% additional masking
 ($\dagger$), averaging losses over two draws.
 GENERATORv2 uses temperature calibration fitted on separate data ($\ddagger$).
 B: additional masking substantially improves post-trained NTv3 sequence loss.
 C: calibration improves GENERATORv2's predicted probabilities
 without changing base-prediction accuracy. Whiskers show 95\% block-bootstrap intervals. Colors denote model/baseline category.}
 \label{fig:prior}
\end{figure}

\paragraph{The models learn useful sequence priors.} To compare causal and masked sequence models, we mask out the central base of 512 GRCh38 windows of 1,200bp. For this base reconstruction, NTv3 650M post and Evo2 have the lowest losses among models in Figure~\ref{fig:prior}A. The foundation models outperform simple base composition statistics, but do not seem to have converged on prediction accuracy. For all NTv3 models, we used a fixed
12\% additional masking from its training objective, and the final scores averaged the central-base loss over two draws.

\paragraph{Learned sequence knowledge is not always readily accessible.}
For NTv3 650M post, hiding only the central base gives a loss of $2.199$ bits. Masking an additional 12\% reduces this to $1.375$, the lowest among tested models, although heavier masking eventually worsens performance (Figure~\ref{fig:prior}B). Averaging predicted probabilities across eight masking draws reduces loss further, to $1.346$. We tested whether a fixed shift in internal activations could reproduce the benefit of additional masking. We estimated activation differences between single-mask and extra-masked queries, selected layer and shift strength on separate data, and evaluated the shifted model on single-mask queries from held-out genomic blocks. For NTv3 650M post, accuracy rose from 29\% to 56\%, close to the pre-trained model's 55\% without changing weights. Full results, controls and uncertainty estimates are provided in Appendix~\hyperref[app:prior-mode-decision]{\ref*{app:prior-routing}}.

A simpler adjustment substantially improves GENERATORv2: temperature calibration lowers loss from $2.168$ to $1.711$ while leaving accuracy at $50.2\%$ (Figure~\ref{fig:prior}C). Together, these results show that measured sequence-prior performance depends on how the model is queried and its outputs are used.

These gains do not establish that sequence priors capture functional effects or variation between individuals. Evo2 and CENO reconstruct $55.2/62.4\%$ of assay reference bases, yet their allele scores correlate only $0.045/0.029$ with reporter effects (Appendix~\ref{app:prior-functional}). Our main prior evaluation uses GRCh38; extending strong sequence priors to natural haplotypes remains an important next step (evident by results in Appendix~\ref{app:prior-memory}).

\section{\capheading{FunctionalPrediction}{from sequence to experimental response}} \label{sec:functional-prediction}

\begin{figure}[!tb]
 \setlength{\parskip}{0pt}
 \centering\capfigure{FunctionalPrediction}
 \includegraphics[trim=0bp 6.8bp 0bp 2bp,clip,width=0.90\linewidth]{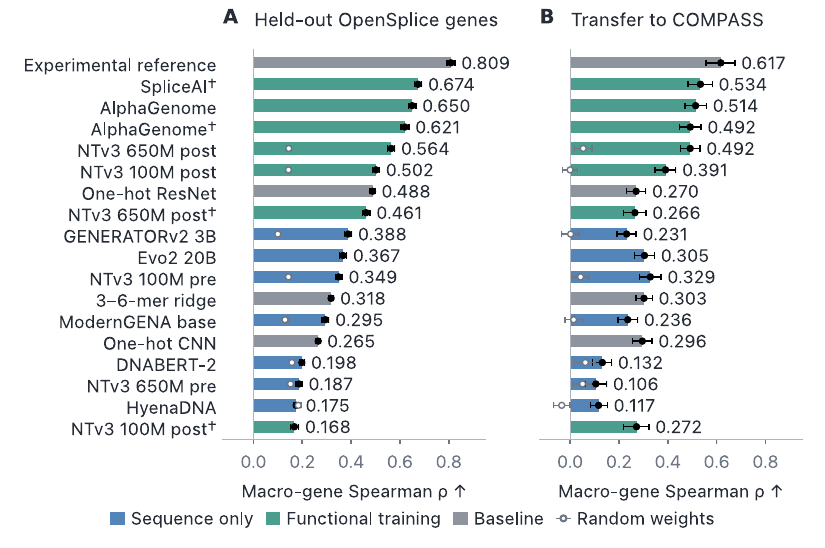}
 \capcaption{FunctionalPrediction}{\textbf{Functional post-training supports transfer across assays and genes.}
A: mutation-induced changes in exon inclusion on
OpenSplice, with complete genes held out of readout fitting
(466 genes).
B: transfer of OpenSplice-trained predictors to 4,618 COMPASS
variants in 88 genes excluded from source training, without
COMPASS fitting or model selection.
Scores average within-gene Spearman correlations.
Experimental references compare repeat measurements. Native heads ($\dagger$) have splicing-related outputs. Gray hollow points show random network weights while whiskers are
 95\% gene-bootstrap intervals (Appendix~\ref{app:simulate}).}
 \label{fig:simulate}
\end{figure}

\paragraph{Mutation-effect prediction transfers.}

Next, we explore whether genomic models can predict how mutations affect RNA splicing. OpenSplice measures changes in exon inclusion across 590,104 mutations in 466 genes \citep{quarantani2026opensplicepaper}. On held-out genes, AlphaGenome's embedding-fitted linear head reaches macro-gene Spearman 0.650, compared with 0.621 for its native outputs and 0.564 for NTv3 650M post (Figure~\ref{fig:simulate}A). An older specialist model SpliceAI \citep{jaganathan2019spliceai} reaches 0.674, higher than the top-performing foundation models.

NTv3 enables a comparison of the same architecture before and after functional training (post-training). Post-training improves mutation-effect prediction at both model sizes, and the readout from the 650M post-trained model significantly outperforms both a readout from the matched architecture with random weights (0.144) and a task-trained one-hot ResNet (0.488). Efficiency of training labels and direction analyses are found in Appendix~\ref{app:simulate}.

We then test whether the OpenSplice-trained readouts transfer to a different assay. Readouts trained on OpenSplice are evaluated on 4,618 COMPASS mutations across 88 genes excluded from the OpenSplice training set, without using COMPASS labels for fitting or selection \citep{koplik2025compass}. Both AlphaGenome and NTv3 650M post retain substantial predictive accuracy for mutation effects averaged across cell lines, outperforming the k-mer baseline fitted on the OpenSplice training set (Figure~\ref{fig:simulate}B). Their performance is similar to that of native SpliceAI and AlphaGenome. These results suggest that the models capture patterns that remain useful beyond both the experimental assays and genes used for fitting.

\paragraph{Editing efficiency depends on genomic context.}

The K562 TRIP library tests editing efficiency across genomic locations while holding the reporter and editing program fixed
\citep{mathis2025pridict2}. Both post-trained NTv3 models improve over
their pre-trained counterparts, while their differences from AlphaGenome are not statistically resolved (Appendix~\ref{app:editing-context}).

\section{\capheading{BiologicalContext}{how does function depend on the context}}
\label{sec:biological-context}

\begin{figure}[!tb]
 \setlength{\parskip}{0pt}
 \centering\capfigure{BiologicalContext}
 \includegraphics[trim=0bp 0.7bp 0bp 5bp,clip,width=.97\linewidth]{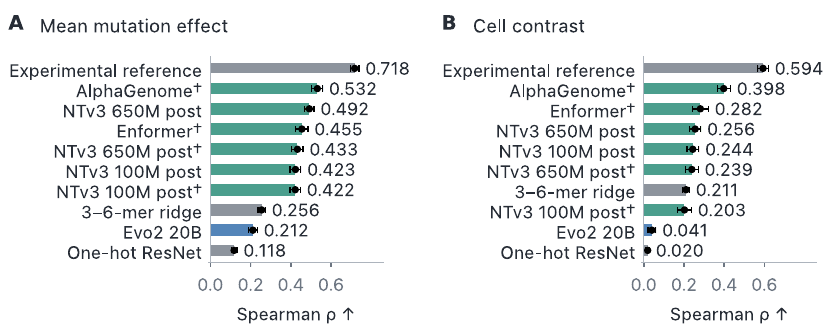}
\capcaption{BiologicalContext}{\textbf{Predicting cell-specific effects.}
A/B: mean mutation effect and HepG2--K562 difference across
128 Siraj regulatory elements.
Experimental references compare groups of repeat measurements (127 elements);
they are not ceilings.
$\dagger$ marks native outputs; unmarked NTv3 entries use fitted readouts.
Whiskers show available 95\% element-bootstrap intervals.}
\label{fig:context}
\end{figure}

\paragraph{Mutation effects differ between cells.}
Models may easily predict a mutation's average effect yet struggle to predict such effect when conditioned on the biological context. Commonly, one may be interested in specific tissues or cell types. %
In the Siraj dataset, the same mutations are measured in both HepG2 and K562 cells, allowing us to evaluate both their mean effect and the difference between cell types
\citep{siraj2026functional}. Across 128 regulatory elements, AlphaGenome predicts the mean effect with a correlation of 0.532, compared with 0.398 for the difference between them (Figure~\ref{fig:context}A/B). It uses its existing prediction tracks without fitting to Siraj measurements. Linear readouts fitted to NTv3 650M post reach 0.492 and 0.256, respectively. Both models predict the mean
more accurately than the difference, with a larger gap for NTv3.

\paragraph{Expression profiles distinguish tissues.}

We also evaluate how well models distinguish expression profiles across tissues, using 5,782 held-out genes measured in 51 GTEx tissues \citep{gtex2020atlas}. {For each gene, we pair each tissue with the tissue where measured expression differs most. We then measure how well a model's difference in expression prediction aligns with the observed difference. Using native outputs, AlphaGenome reaches a correlation of 0.775, compared with 0.594 for NTv3 650M post (Figure~\ref{fig:context-tissues}A). Because the comparison tissue is chosen using measured expression, this effectively tests model predictions for observed tissue contrasts.}

A second test asks whether a predicted expression profile across genes identifies the correct tissue among the 51 candidates. Using native outputs, AlphaGenome achieves 68.6\% accuracy, compared with 45.1\% for NTv3 650M post and merely 7.8\% for the k-mer ridge (Figure~\ref{fig:context-tissues}B). Random chance has 2.0\% accuracy, while the experimental reference identifies tissues with 99.0\% accuracy. This reference uses expression profiles from independent donor halves, averaging both directions. The models generally allow for accurate tissue classification, but a sizeable performance gap remains. The evaluations hold out genes within each tissue (see also Appendix~\ref{app:gtex-tissue-contrast}).

\section{\capheading{GeneticInteractions}{predicting departures from additivity}}
\label{sec:genetic-interactions}

\begin{figure}[!tb]
 \setlength{\parskip}{0pt}
 \centering\capfigure{GeneticInteractions}
 \includegraphics[trim=0bp 5.4bp 0bp 2.5bp,clip,width=0.90\linewidth]{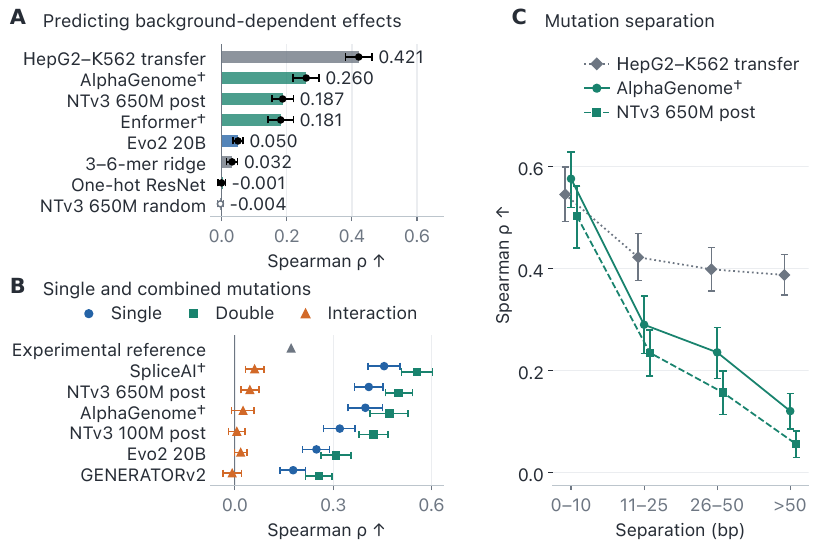}
\capcaption{GeneticInteractions}{\textbf{Accurate combined effects
can conceal weak interaction prediction.}
A: difference in mutation effects between two backgrounds differing at one base.
B: single, double and non-additive mutation effects in 110 COMPASS genes.
C: Interaction prediction by distance between mutations.
HepG2--K562 transfer compares measured interactions across cells (A/C).
The experimental reference compares repeat measurements of interactions (B).
Neither is a ceiling.
$\dagger$ marks native outputs, and whiskers show 95\% bootstrap intervals over regulatory elements (A/C) or genes (B).}
\label{fig:compose}
\vspace{6pt}
\end{figure}

\paragraph{A mutation's effect depends on its genetic background.}
Rather than the cell level context of the previous section, different genetic backgrounds may also alter the effect of mutations. The Siraj data allows for isolating this interaction, since it contains the same edit on two paired allelic backgrounds. Across 68,026 paired mutations in 114 regulatory elements, AlphaGenome reaches interaction correlation 0.260, compared with 0.187 for NTv3 650M post's fitted readout and 0.181 for Enformer \citep{avsec2021enformer} (Figure~\ref{fig:compose}A). Predictions are most accurate for mutations within 10 bp of one another and become less accurate with increasing separation (Figure~\ref{fig:compose}C).

\paragraph{Combined-effect accuracy can conceal weak interaction prediction.}

COMPASS tests non-additive interactions on splicing by measuring single and double mutations relative to the same reference. On its 110-gene panel, NTv3 650M post reaches correlation 0.409 for single-mutation effects and 0.500 for double-mutation effects, but only 0.046 for the interaction remaining after subtracting the two single effects (Figure~\ref{fig:compose}B). Native splice predictors show the same pattern, and replicate correlation for the interaction measurements is 0.171.

To assess how much interaction prediction contributes to overall accuracy, we remove the predicted interaction and retain only the additive effects. NTv3's double-mutation correlation changes only from 0.500 to 0.493. Similarly, removing AlphaGenome's predicted background dependence on Siraj changes overall correlation from 0.528 to 0.521, a small but statistically supported decrease
of 0.0070 (95\% interval: 0.0014--0.0120;
Appendix~\ref{app:interaction-ablation}). This indicates that a model can predict combined effects accurately while recovering little of the interaction between mutations.

\section{\capheading{GenomicDesign}{selecting sequences for a desired response}} \label{sec:genomic-design}

\begin{figure}[!tb]
 \setlength{\parskip}{0pt}
 \centering\capfigure{GenomicDesign}
 \includegraphics[trim=0bp 4.6bp 0bp 3bp,clip,width=0.90\linewidth]{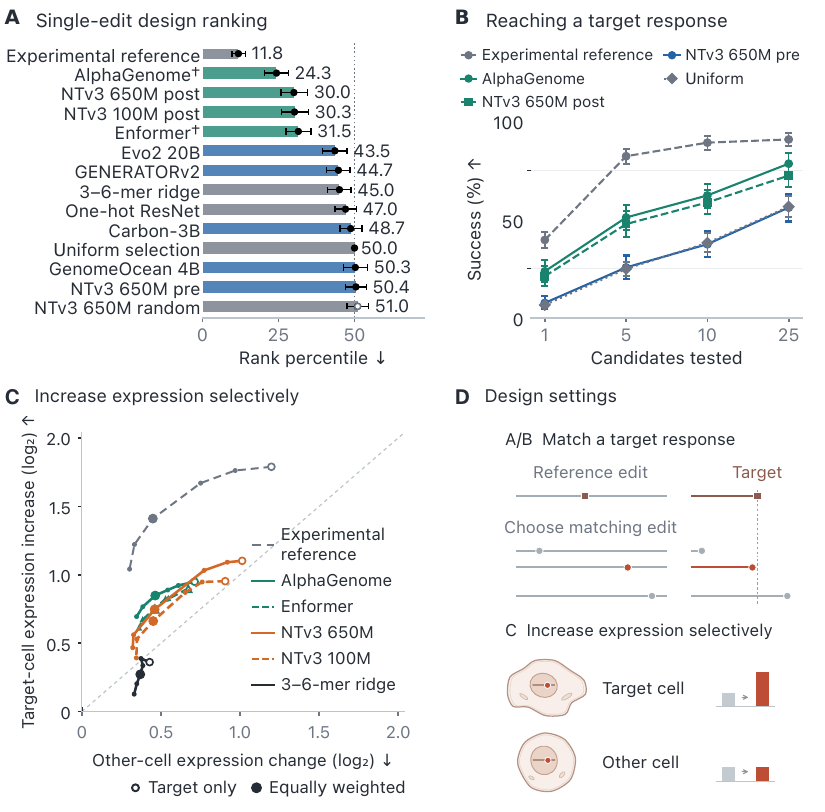}
\capcaption{GenomicDesign}{\textbf{Selecting edits for a desired
response and cell specificity.}
A/B: rank of the first selected edit and success at matching a
reference-edit response as testing budget increases
(331 targets, 110 regulatory elements).
C: target-cell expression increase versus absolute other-cell
change across 127 loci, varying the penalty on unwanted effects.
Open markers ignore the other cell ($\lambda=0$), and filled markers penalize it equally ($\lambda=1$). Experimental references in A--C select edits on one group of repeat
measurements and evaluate on the other. Models may outperform these references.
D: task settings.
$\dagger$ marks native outputs in A, and whiskers show 95\% element-bootstrap intervals.} \label{fig:control}
\end{figure}

We finally ask whether model predictions help choose DNA with a desired response. These tasks evaluate the measured outcomes of model-selected candidates, progressing from matching a single response to controlling expression across multiple cell types. All choices are made within experimentally assayed libraries, with candidate measurements withheld from selection.

\paragraph{Matching a target response.}

Can a model find an alternative mutation that reproduces the effect of a reference edit? In the Siraj assay, each element contains a natural variant and a measured change in activity, which we define as the target response. Excluding that position leaves 578–597 candidate substitutions per element. A model predicts an effect for every candidate and the reference edit, then ranks candidates by how closely their predicted effects match its own reference prediction. AlphaGenome scores effects as changes in its cell-matched accessibility outputs; NTv3 uses linear heads evaluated on held-out elements. All candidates' measured effects are withheld until scoring.

For evaluation, we rank edits by how closely their measured effects match the target. Across 331 targets in 110 elements, AlphaGenome's first choice has an average percentile rank of 24.3, compared with 30.0 for NTv3 650M post and 50 for uniform selection (Figure~\ref{fig:control}A; lower is better). The top ten choices include at least one edit that matches the reference response within assay uncertainty for 62\% of targets with AlphaGenome and 59\% with NTv3, compared with 38\% under uniform selection (Figure~\ref{fig:control}B). Model rankings can therefore help find matching edits with fewer tests.

\paragraph{Changing expression selectively.}
\label{sec:cell-selective}

We also test a more challenging task, choosing edits which increase expression in one cell type while minimally changing expression in another. We evaluate 167,958 substitutions across 280 parental backgrounds at 127 Siraj loci, with HepG2 and K562 alternately serving as the target cell. Each model chooses one substitution or no edit by maximizing predicted utility $U_{\mathrm{cell}}=t-\lambda|o|$, where $t$ and $o$ are expression changes relative to the unmutated parent in the target and other cell. Six fixed penalties describe the tradeoff between the two responses (Figure~\ref{fig:control}C). Complete loci were held out of readout fitting and native-output calibration.

At equal weighting ($\lambda=1$), calibrated native AlphaGenome achieves measured utility 0.384, compared with 0.287 for NTv3 650M post, 0.211 for NTv3 100M post and zero for no edit. All three models improve over controls that choose edits at the same rate and with similar predicted effects in the target cell. AlphaGenome's utility is 0.086 higher than its matched control (adjusted 95\% interval: 0.030--0.142). With a linear readout fitted to its frozen embeddings, AlphaGenome reaches a utility of 0.515 (Table~\ref{tab:bend2-reference}).

To see whether better choices are possible among the substitutions tested in the Siraj assay, we select edits using their measured effects in HepG2 and K562. We use one group of repeat measurements to choose edits and the other to evaluate them, then swap the groups and average the results. On these same evaluation subsets, experimental selection outperforms native AlphaGenome by 0.603 in utility, with higher utility at 108 of 127 loci. This reference demonstrates reproducible opportunities for better selection within the library (Appendix~\ref{app:cell-selective}). Similarly, we test genomic design selective for the specific allelic background in Appendix~\ref{app:final-siraj}.

\paragraph{Selecting multi-cell expression programs.}
\label{sec:multicell-programming}

The final task is to select for activity in several cell types while avoiding activity elsewhere. From the twelve-cell Atlas data
\citep{castillohair2025atlas}, we aim for sixteen expression
patterns: eight with two cell types ON and ten OFF, and eight with three ON and nine OFF. For each pattern, a model needs to choose ten enhancers from a sequence bank of 2,411 enhancers (each 145-nt long). Models therefore need to rank them using predicted activities in the score
\begin{equation}
 U_{\mathrm{multi}}(x)
 =
 \min_{c\in\mathrm{ON}}y(x,c)
 -
 \max\!\left(0,\max_{c\in\mathrm{OFF}}y(x,c)\right),
 \label{eq:multicell-utility}
\end{equation}
where $y$ is $\log_2$ reporter activity relative to negative controls. Our score rewards activity in the weakest ON cell and penalizes activity above the control level in the strongest OFF cell. Readouts are fitted on 5,246 other sequences where all twelve possible cell types are represented. We additionally exclude sequence families shared with the candidates.

\begin{figure}[!t]
 \setlength{\parskip}{0pt}
 \centering\capfigure{GenomicDesign}
 \includegraphics[trim=0bp 0bp 0bp 5.5bp,clip,width=0.9\linewidth]{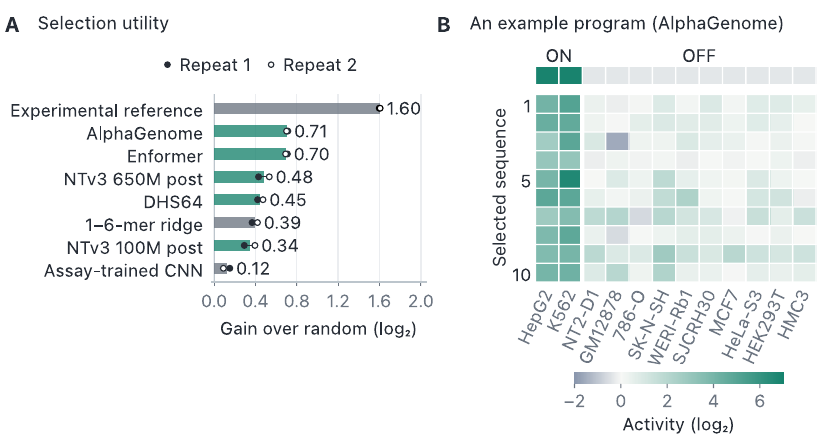}
\capcaption{GenomicDesign}{\textbf{Selecting DNA with activity
in desired cells and little activity elsewhere.}
Models choose ten enhancers for each of sixteen requested expression patterns.
A: gain in utility over random selection among sequences originally
designed for the same ON/OFF pattern, averaged across patterns.
Utility rewards the weakest ON-cell activity and penalizes the
strongest OFF-cell activity above the negative-control level.
Dots show the two repeat measurements.
The experimental reference selects sequences using one measurement
and evaluates them using the other, then swaps them.
B: example across cell types, showing ten enhancers chosen by AlphaGenome having selective activity in HepG2 and K562.} \label{fig:frontier-programming}
\end{figure}

We compare the models' ranked choices with enhancers originally designed by Atlas to be active in the target cell types and inactive elsewhere. With linear readouts fitted to native outputs, the gains in utility are 0.707 for AlphaGenome, 0.700 for Enformer, 0.482 for NTv3 650M post and 0.344 for NTv3 100M post (Figure~\ref{fig:frontier-programming}A). For all four models, accounting for OFF-cell activity improves selection over using ON-cell activity alone, across all sixteen patterns and both repeat measurements (Appendix~\ref{app:multicell-programming}).

For this multi-cell task, however, average utility remains negative: $-0.222$ for AlphaGenome and $-0.228$ for Enformer. A linear readout fitted to AlphaGenome's frozen embeddings raises the gain over random selection to 0.833, but mean utility is still negative ($-0.096$; Table~\ref{tab:bend2-reference}). By comparison, choosing sequences using one repeat measurement and evaluating them using the other gives a mean utility of $+0.673$. Better matches to the desired expression patterns are therefore available among the tested sequences.

\section{Conclusions}

Genomic models capture functional information that transfers across datasets and genes. This genomic understanding helps them select DNA with a desired target response. Functional training largely expands these capabilities: predictions from embedding-fitted linear heads also match or exceed native outputs, suggesting that its benefits extend beyond the prediction heads and affect the models more deeply. At the same time, accurate aggregate predictions can conceal weak understanding of cell types and genetic interactions, which on the other hand are highly important for practical downstream use. BEND 2.0 focuses on the capabilities that matter for genomic design, hoping to support the development of models that make this search more effective.

Tasks with less consistent experimental results are often the more difficult tasks for the models, meaning performance should be interpreted in light of the experiment's reliability. The chosen experimental references are somewhat naive (for example, transferring mutation effects across cell types), yet they still beat current models. The linear heads may also depend on the embedding sizes of the models, and models with different input context lengths may have an effect on task performance. Despite these limits, BEND 2.0 shows where genomic models may help narrow hundreds of candidate mutations to a testable few, and where further progress is needed.

\subsection*{Acknowledgements}
This work was supported by the Novo Nordisk Foundation through the Center for Basic Machine Learning Research in Life Science (NNF20OC0062606), CAZAI (NNF22OC0077058), Quantum for Life (NNF20OC0059939), and the NovoSTAR programme, and by the Pioneer Centre for AI (DNRF grant P1). We also acknowledge the Danish Centre for AI Innovation (DCAI) and the Novo Nordisk Foundation (NNF25OC0105165) for supporting our access to the Gefion supercomputer.

\subsection*{AI use statement}
Generative AI tools were used in evaluating models on HPC clusters, preparing code, organizing the tasks, and preparing the manuscript and supporting materials. The authors reviewed AI-assisted work and take full responsibility for the final manuscript, including its methods, results and conclusions.

\subsection*{Ethics statement}
This work evaluates machine learning models using existing functional-genomics experiments, reference sequences and population-genomic resources. {Human genomic data can be identifying. We used the data in accordance with the applicable access and redistribution conditions, which also govern future use. Performance on the contexts presented here} should not be assumed to generalize uniformly to other populations or biological settings. Models supporting genomic design may also have dual-use applications. Downstream use therefore requires appropriate biological validation, biosafety review and oversight.

\subsection*{Reproducibility statement}
We provide all code and fixed data splits for running BEND 2.0. We also host the benchmarking data in our own
\href{https://huggingface.co/datasets/yhsure/BEND2}{Hugging Face repository}
so future evaluations do not depend on the original hosts, making our benchmark more sturdy. Data are redistributed under their original terms, with attribution and licensing information included in the repository. Our code includes both setup and evaluation instructions, while our appendices contain additional methodological details and supporting results.

\bibliography{references}
\bibliographystyle{iclr2027/iclr2027_conference}

\appendix
\clearpage
\begingroup
\pdfbookmark[0]{Appendix contents}{appendix-contents}
\hypersetup{hidelinks}
\color{black!88}
\renewcommand{\capheading}[2]{\caplabel{#1}: #2}
\setlength{\parindent}{0pt}
\setlength{\parskip}{0pt}
\fontsize{10}{14.5}\selectfont

{\normalfont\scshape\fontsize{14}{19}\selectfont\color{black}%
 \textls[20]{Appendix contents}\par}
\vspace{16pt}

\newcommand{\appcontentsline}[3]{%
  \par\begingroup
  \leftskip=#1\relax
  \advance\leftskip by #2\relax
  \rightskip=22pt
  \parfillskip=-\rightskip
  \leavevmode\hskip-#2\relax
  \hyperref[#3]{\makebox[#2][l]{\ref*{#3}}\nameref*{#3}}\nobreak
  \hspace{5pt}{\normalfont\color{black!24}\leaders\hbox{\kern2pt.\kern2pt}\hfill}\nobreak
  \makebox[22pt][r]{\normalfont\color{black!65}\hyperref[#3]{\pageref*{#3}}}\par
  \endgroup}
\newenvironment{appcontentsgroup}[1]{%
  \par\addvspace{13pt}
  {\bfseries\appcontentsline{0pt}{20pt}{#1}}%
  \vspace{3pt}
}{\par}
\newcommand{\appcontentsentry}[1]{%
  \appcontentsline{20pt}{28pt}{#1}}

\begin{appcontentsgroup}{app:evaluation}
  \appcontentsentry{app:predictions-readouts}
\end{appcontentsgroup}

\begin{appcontentsgroup}{app:prior}
  \appcontentsentry{app:prior-routing}
  \appcontentsentry{app:prior-repair}
  \appcontentsentry{app:prior-memory}
  \appcontentsentry{app:prior-functional}
\end{appcontentsgroup}

\begin{appcontentsgroup}{app:simulate}
  \appcontentsentry{app:editing-context}
\end{appcontentsgroup}

\begin{appcontentsgroup}{app:context}
  \appcontentsentry{app:gtex-tissue-contrast}
  \appcontentsentry{app:personal-native}
\end{appcontentsgroup}

\begin{appcontentsgroup}{app:genetic-interactions}
  \appcontentsentry{app:compass-predictions}
  \appcontentsentry{app:interaction-ablation}
  \appcontentsentry{app:conditional-development}
\end{appcontentsgroup}

\begin{appcontentsgroup}{app:sequence-choice}
  \appcontentsentry{app:response-matching}
  \appcontentsentry{app:final-siraj}
  \appcontentsentry{app:cell-selective}
  \appcontentsentry{app:multicell-programming}
\end{appcontentsgroup}

\begin{appcontentsgroup}{app:statistical-comparisons}
  \appcontentsentry{app:confidence-intervals}
\end{appcontentsgroup}
\endgroup
\clearpage

\section{Evaluation details}
\label{app:evaluation}

Holding out data for these experiments does not imply that the sequences were absent from foundation-model pretraining.

\subsection{Predictions and readouts} \label{app:predictions-readouts}
\paragraph{Native outputs and fitted linear heads.}
We evaluate models using native outputs and embedding-fitted linear heads. Native outputs are produced by the models' existing output heads. Depending on the model, these include self-supervised sequence logits, supervised predictions of functional measurements or genomic annotations. Embedding-fitted linear heads are trained on sequence embeddings while the original model is kept frozen. For multi-cell design, we also fit linear heads to native outputs; these are identified separately. Any task-specific calibration of native outputs is described with the task.

The use of native outputs and fitted linear heads varies depending on the task. Below are examples given for how we obtain variant-effect scores with the different approaches.

To use native outputs, where sequence output logits are available, we score a variant in causal models as its change in log sequence probability. For masked models, the score is calculated as the alternate-versus-reference log odds:
\begin{equation}
  \begin{aligned}
  s_\theta^{\mathrm{causal}}(\delta\mid x)
  &=\log p_\theta(x+\delta)-\log p_\theta(x),\\
  s_\theta^{\mathrm{masked}}(\delta\mid x)
  &=\log q_\theta(a_{\mathrm{alt}}\mid x_{\setminus i})
   -\log q_\theta(a_{\mathrm{ref}}\mid x_{\setminus i}).
  \end{aligned}
  \label{eq:nativescore-app}
\end{equation}

For AlphaGenome, Enformer and post-trained NTv3, we use their native functional predictions, such as DNase accessibility. For each output $c$, we predict the variant's effect by subtracting the reference prediction from the variant prediction, $f_\theta(x+\delta,c)-f_\theta(x,c)$.

To measure variant effect using an embedding-fitted linear head, we determine the difference between frozen embeddings. With $\phi_\theta(x)$ representing the frozen pooled representation of a sequence and $\Delta_i$ as the measured effect of variant $\delta_i$, we fit ridge regression on the representation change:
\begin{equation}
  w^*=\argmin_w\sum_{i\in\mathrm{train}}
  \left(\Delta_i-w^\top\Delta\phi_\theta(\delta_i\mid x_i)\right)^2
  +\lambda\lVert w\rVert_2^2,
  \qquad
  \Delta\phi_\theta(\delta_i\mid x_i)
  =\phi_\theta(x_i+\delta_i)-\phi_\theta(x_i).
  \label{eq:ridge-app}
\end{equation}

Feature standardization is estimated on the training data only. Projections used to reduce feature dimensionality are random, seeded and label-independent, and the ridge penalty $\lambda$ is either fixed in advance or chosen on validation data. Per-task settings are provided in their corresponding sections.

Embedding-fitted linear heads evaluated on COMPASS are fit only on OpenSplice.

\paragraph[NTv3 outputs.]{\phantomsection\label{app:ntv3-interfaces}NTv3 outputs.} Because NTv3 produces multiple native outputs, we define here the native outputs used for each figure. Post-trained NTv3 models produce nucleotide logits, functional tracks and genome annotations \citep{boshar2025ntv3}. For example, native-output results use nucleotide logits for the sequence-prior task, predicted splice-boundary probabilities for splicing (Figure~\ref{fig:simulate}), and cell-matched DNase outputs for cell contrasts (Figure~\ref{fig:context}). These evaluations use the existing output heads. NTv3 also uses embedding-fitted linear heads and, for multi-cell design, linear heads fitted to native outputs. Throughout the work, we specify the exact settings during each task.

The pre-trained NTv3 100M and 650M models produce nucleotide logits only and have no functional heads.

\paragraph{AlphaGenome fitted readouts.}
AlphaGenome's supported input is larger than the benchmark's sequences, so padding is used. We take frozen 1-bp representations from the final model layer, mean-pool over the inserted bases, project to 512 features, and fit ridge with $\alpha=10$. Siraj and OpenSplice/COMPASS inserts use 16,384-bp N-padded inputs without genomic or vector flanks. We hold out complete elements for Siraj. The COMPASS transfer test in Figure~\ref{fig:simulate}B excludes overlapping sequences and removes the 23 shared genes from both datasets. For GTEx, we use 131-kb inputs, scale features using only the training data, and choose $\alpha=100$ using validation genes. The other design readouts follow Appendices~\ref{app:cell-selective} and \ref{app:multicell-programming}. Native and fitted results remain separate.

\paragraph[Experimental references.]{\phantomsection\label{app:experimental-references}Experimental references.}
For experimental references, we use one set of measurements to make predictions or choose sequences, then evaluate these predictions or choices using another set. Depending on the setting, the references can be somewhat naive (for example, transferring effects across cell types) or stronger (predicting the same response using repeat measurements).

The split is dependent on the experimental data:
\begin{itemize}
    \item For OpenSplice, we compare the average of two variant PSI repeats with the third, averaging all three rotations. Effects share the reported wild-type estimate.
    \item For Siraj, the background-interaction reference compares measurements between
HepG2 and K562. Other references compare effects calculated separately from repeats 1--2 and 3--5. We evaluate model predictions against effects estimated from all repeats. For cell-selective design, we instead evaluate both model choices and experimental references using effects relative to the unmutated parent.
    \item For GTEx, tissue identification averages both
    donor-half directions (99.0\%), and its uncertainty was not re-estimated.
    \item For Atlas, each of the two RNA measurements is
evaluated against the other.
\end{itemize}

For the Siraj data, the references for cell effects use 127 of the 128 Siraj elements, as one lacks the required count data. Ranking and target-response matching use all 331 targets in 110 elements. We select on one repeat group, test on the other, then swap the groups and average the results.

\paragraph{Uncertainty and scoring.}
We estimate uncertainty by bootstrapping whole biological groups, such as genes, regulatory elements or loci, following the grouping used for the data splits. This makes measurements sharing a unit stay correlated. When comparing a pre-trained model with a randomly initialized network of the same architecture, we evaluate both on the same bootstrap resample. Experiments with several training seeds jointly resample seeds and biological units. We calculate Spearman correlations within each group and average them, giving every group equal weight:
\begin{equation}
  \rho_{\G}(\hat y,y)=\frac{1}{|\G|}\sum_{g\in\G}
  \operatorname{Spearman}(\hat y_g,y_g).
  \label{eq:macro-app}
\end{equation}

\section{\capheading{SequencePriors}{Sequence probability and reconstruction}}
\label{app:prior}

\paragraph[Sampling and controls.]{\phantomsection\label{app:sampling_control}Sampling and controls.}
We sample 512 autosomal GRCh38 windows, each 1,200 bp long and from distinct 1-Mb blocks. Positions and strands are chosen randomly. Evaluation blocks are separate from those used to fit sequence baselines and calibrate model probabilities.

Uniform DNA costs two bits per base.  The tuned Markov model uses order ten and smoothing strength 64, chosen on a 768/256 split and then refitted on all 1,024 fitting windows. Confidence intervals use 4,000 paired resamples of complete genomic blocks. Full settings and seeds are preserved in our codebase.

\paragraph{Sequence probability.}
For the causal models, we use
\begin{equation}
  H_\theta=-\frac{1}{600}\log_2 p_\theta(x_{601:1200}\mid x_{1:600}).
  \label{eq:sequence-density-app}
\end{equation}
Evo2 20B and CENO use base tokens \citep{brixi2026evo2,ma2026ceno}. Their probabilities are normalized over all tokens, including non-DNA tokens. For GENERATORv2, we score each sixmer by multiplying the base probabilities at its six positions \citep{li2026generator2}.

\begin{figure}[ht]
  \centering
  \capfigure{SequencePriors}
  \includegraphics[width=.98\linewidth]{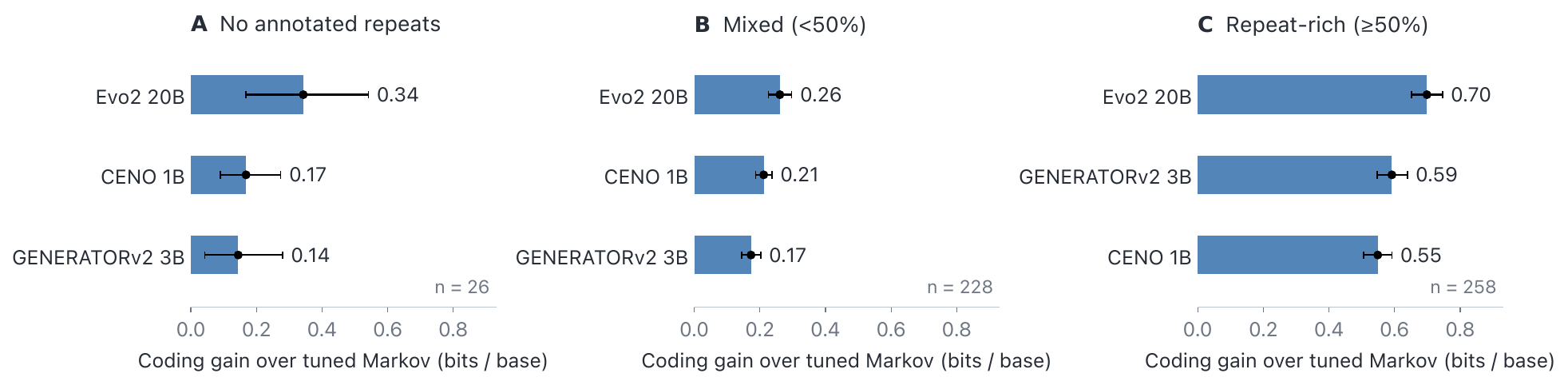}
  \capcaption{SequencePriors}{\caplead{SequencePriors}{Sequence prediction across repeat content.}
  Reduction in base loss relative to the tuned Markov model.
  Higher is better. Black bars show paired 95\% bootstrap confidence
  intervals; numbers give window counts.}
  \label{fig:prior-repeats}
\end{figure}

\paragraph{Sequence prediction in repetitive DNA.}
First, we group windows in three categories based on the count of bases annotated as repeats in GRCh38: none, some but less than half, or at least half. The three causal models have lower base loss than the tuned Markov model in every group, with the largest gains in repeat-rich windows (Figure~\ref{fig:prior-repeats}).

\paragraph{Reconstructing a hidden base.}
We hide base 601 in each window. For each causal model, we insert A, C, G or T and score the entire resulting sequence, including the bases after the changed position. We normalize the four sequence probabilities to obtain a probability for each candidate base:
\begin{equation}
  q_\theta(a\mid x_{\setminus i})=
  \frac{p_\theta(x^{(a)})}
  {\sum_{b\in\{\mathrm{A,C,G,T}\}}p_\theta(x^{(b)})}.
  \label{eq:base-recovery-app}
\end{equation}
NTv3 predicts the masked base directly \citep{boshar2025ntv3}. For the main comparison, we average the base loss over two draws, each masking an additional 12\% of the other positions. We also compare a single-mask query with the average base probabilities from eight such draws. This masking rate comes from NTv3's training procedure. Post-trained NTv3 uses the human species setting.

We report base log loss $-\log_2 q(x_i\mid x_{\setminus i})$ and accuracy.

\subsection{How post-training changes access to the sequence prior}
\label{app:prior-routing}

Additional masking improves base prediction in post-trained NTv3 despite --- somewhat counterintuitively --- removing more of the surrounding DNA context (Figure~\ref{fig:prior}B). We investigate this effect and test whether the improved performance can be reproduced while only masking the target base.

\paragraph{Mask placement.}
We test both NTv3 sizes before and after post-training on the 512 benchmark windows. Each query masks the target base and alters 144 other positions, using two random draws. Masks are spread across the window, placed within 128 bases of the target, or kept more than 256 bases away. For distant masks, we compare scattered and compact patterns with the same number of masks on each side and matched average distances. We also replace the same distant bases with other nucleotides instead of MASK tokens.

\begin{figure}[ht]
  \centering
  \capfigure{SequencePriors}
  \includegraphics[width=.98\linewidth]{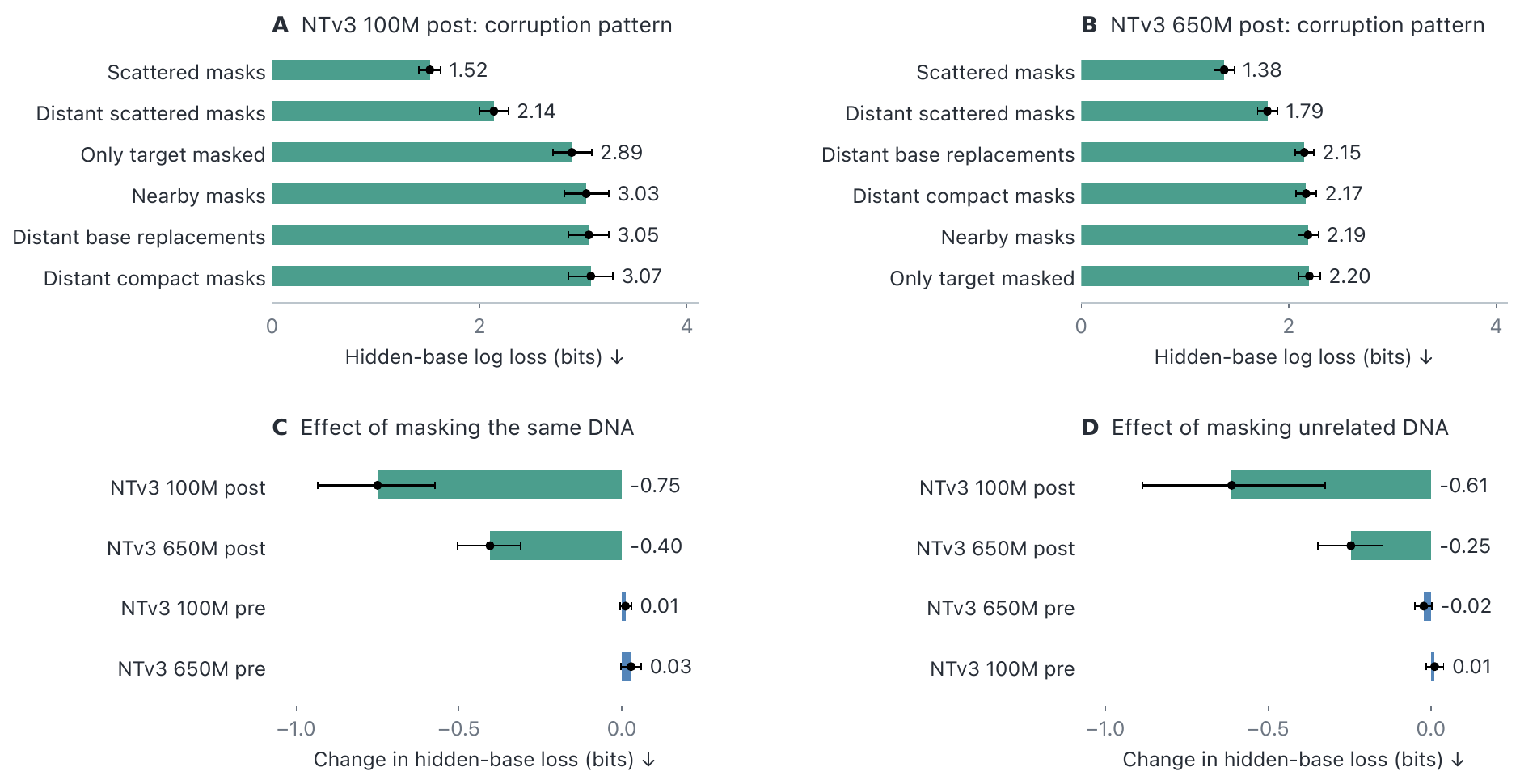}
  \capcaption{SequencePriors}{\caplead{SequencePriors}{Base prediction depends on the masking pattern.}
  \emph{A--B:} Base loss with 144 additional masked or replaced bases.
  Distant changes leave 256 bases on each side of the target untouched.
  \emph{C--D:} Change in loss when replacing the transformer representation,
  comparing versions with and without extra masks. Representations come
  from the same DNA in C and unrelated DNA with similar GC content in D.
  Skip features stay fixed. Blue denotes pre-trained and teal post-trained
  models. Whiskers show bootstrap intervals over genomic blocks (A--C)
  or sequence pairs (D).}
  \label{fig:prior-routing}
\end{figure}

For 650M post, distant scattered masks give lower loss than compact masks ($1.794$ versus $2.167$ bits; paired difference $0.373$, 95\% interval $0.283$--$0.467$). Mask count and average distance therefore do not explain the difference. Base replacement does not reproduce the benefit. The 100M post model behaves similarly, while pre-trained models change much less (Figure~\ref{fig:prior-routing}A--B).

\paragraph{Replacing transformer representations.}
NTv3 predicts bases using both its final transformer representation and convolutional features that bypass the transformer (skip features)
\citep{boshar2025ntv3}. With only the target base masked, we replace the
transformer representation with one obtained from the same DNA with extra masks or base replacements. We keep the skip features fixed, then test the reverse: replacing only the skip features.

For 650M post, uniform masking lowers loss by $0.821$ bits. Replacing only the transformer representation reproduces $0.812$ bits of this improvement. Replacing only the skip features changes loss by $+0.014$ bits, with an interval that includes zero. Distant masks and the 100M post model show the same pattern (Figure~\ref{fig:prior-routing}C).

\paragraph{Using unrelated DNA.}
We then test whether the masking benefit depends on the sequence that was masked. Using 256 pairs of unrelated sequences with similar GC content, we compare transformer representations from the unrelated DNA with and without extra distant masks. The sequence being predicted and its skip features stay fixed.

Adding masks to the unrelated sequence lowers loss by $0.246$ bits
for 650M post (95\% interval $0.147$--$0.348$) and $0.612$ bits for
100M post ($0.325$--$0.885$). Pre-trained models show no clear improvement (Figure~\ref{fig:prior-routing}D). NTv3 uses clean inputs for functional training and masked or altered inputs for language modeling \citep{boshar2025ntv3}, so masks may help indicate which task to perform.

\paragraph[Fixed activation shift.]{\phantomsection\label{app:prior-mode-decision}Fixed activation shift.}
The benefit from masking unrelated DNA suggests that part of the effect is shared across sequences. We therefore test whether an average change in transformer activations can improve predictions with only the target base masked.

For each model, we use 768 windows from separate genomic blocks. We compare activations with only the target base masked and with 144 additional randomly placed masks. At each transformer block considered, we average the difference across windows and positions:
\begin{equation}
\boldsymbol{\Delta}_{\ell}
=
\frac{1}{NL}
\sum_{n=1}^{N}\sum_{t=1}^{L}
\left[
\mathbf{h}_{\ell,t}\!\left(x_n^{\mathrm{additional}}\right)
-
\mathbf{h}_{\ell,t}\!\left(x_n^{\mathrm{single}}\right)
\right].
\label{eq:activation-difference}
\end{equation}
Here $\mathbf{h}_{\ell,t}(x)$ is the activation at position $t$ of transformer block $\ell$, $L$ is the number of positions, and $N=768$. This gives one fixed vector per model and block.

We choose the block and shift strength on 128 separate windows. The first block with multiplier 1.5 gives the lowest loss for both post-trained sizes, so we use this setting for all four models, each with its own vector. We then test on 1,536 windows from 48 unused genomic blocks, with varying target positions and both strands equally represented. The shift is added at every position, while model weights stay fixed and only the target base is masked.

We compare against reversed and random shifts of the same magnitude, applied at the same block. We also compare with temperature scaling and affine calibration of the four base logits, fitted on the same 768 windows used to estimate the shifts.

The shift raises accuracy from 28.8\% to 51.2\% for 100M post and from
29.1\% to 55.9\% for 650M post, close to the corresponding pre-trained
models (Table~\ref{tab:prior-activation-accuracy}). Pre-trained models change little, and the controls give no comparable improvement. The gains over affine calibration are 20.5 and 25.7 percentage points,
respectively (adjusted 95\% intervals: 15.7--25.5 and 20.9--30.6).
Intervals use block bootstrapping with Bonferroni correction across 28 comparisons.

Much of the drop in base-prediction accuracy after post-training can therefore be recovered through a fixed activation change, without retraining the model.

\begin{table}[!ht]
\centering
\footnotesize
\setlength{\tabcolsep}{5pt}
\captable{SequencePriors}
\begin{tabular}{@{}lcccc@{}}
\toprule
Condition
& \shortstack{NTv3\\100M pre}
& \shortstack{NTv3\\650M pre}
& \shortstack{NTv3\\100M post}
& \shortstack{NTv3\\650M post} \\
\midrule
Unmodified
& 51.3 & 55.3 & 28.8 & 29.1 \\
Temperature scaling
& 51.3 & 55.3 & 28.8 & 29.1 \\
Affine calibration
& \textbf{52.0} & 55.5 & 30.7 & 30.2 \\
Activation shift
& 51.4 & \underline{55.6} & \underline{51.2} & \underline{55.9} \\
Negated shift
& \underline{51.7} & 55.0 & 29.9 & 28.9 \\
Random shift 1
& 51.6 & \textbf{55.8} & 28.9 & 29.8 \\
Random shift 2
& 51.3 & 55.3 & 27.0 & 29.6 \\
Random shift 3
& 51.3 & 55.1 & 30.0 & 29.8 \\
Eight-query average
& 51.5 & 55.1 & \textbf{54.4} & \textbf{58.4} \\
\bottomrule
\end{tabular}
\caprender{SequencePriors}
\caption{\textbf{Masked-base accuracy with activation shifts and probability calibration.}
Accuracy (\%) on 1,536 windows from 48 held-out genomic blocks.
Bold marks the highest accuracy in each column; underlining marks
the second highest. All conditions except the eight-query average
mask only the target base. The eight-query average uses different
additional masks in each query and averages the predicted base
probabilities.}
\label{tab:prior-activation-accuracy}
\end{table}

\subsection{Probability calibration}
\label{app:prior-repair}
Temperature scaling changes confidence but preserves base and allele-score rankings:
\begin{equation}
  q_T(a\mid r)=\frac{q(a\mid r)^{1/T}}{\sum_b q(b\mid r)^{1/T}}.
\end{equation}
We fit one temperature per model on 1,024 separate windows, choosing $0.1\leq T\leq100$ to minimize hidden-base log loss.

\subsection{Sequence prediction on natural haplotypes}
\label{app:prior-memory}

We select 512 nonoverlapping SNP windows on chromosome 22 from the phased 1000 Genomes data for 2,504 participants
\citep{byrska2022highcoverage}. We sample 128 sites in each ALT allele
frequency range: 1--5\%, 20--40\%, 60--80\% and 95--99\%.
Each site then has a REF-carrying sequence, an ALT-carrying sequence and the GRCh38 reference. Carrier sequences include all called variants, including indels. We select windows without using model scores.

\paragraph{Predicting the carried allele.}
We score all four bases in each carrier's own surrounding DNA, giving the two carriers equal weight before averaging across sites.

\begin{figure}[ht]
  \centering
  \capfigure{SequencePriors}
  \includegraphics[width=.98\linewidth]{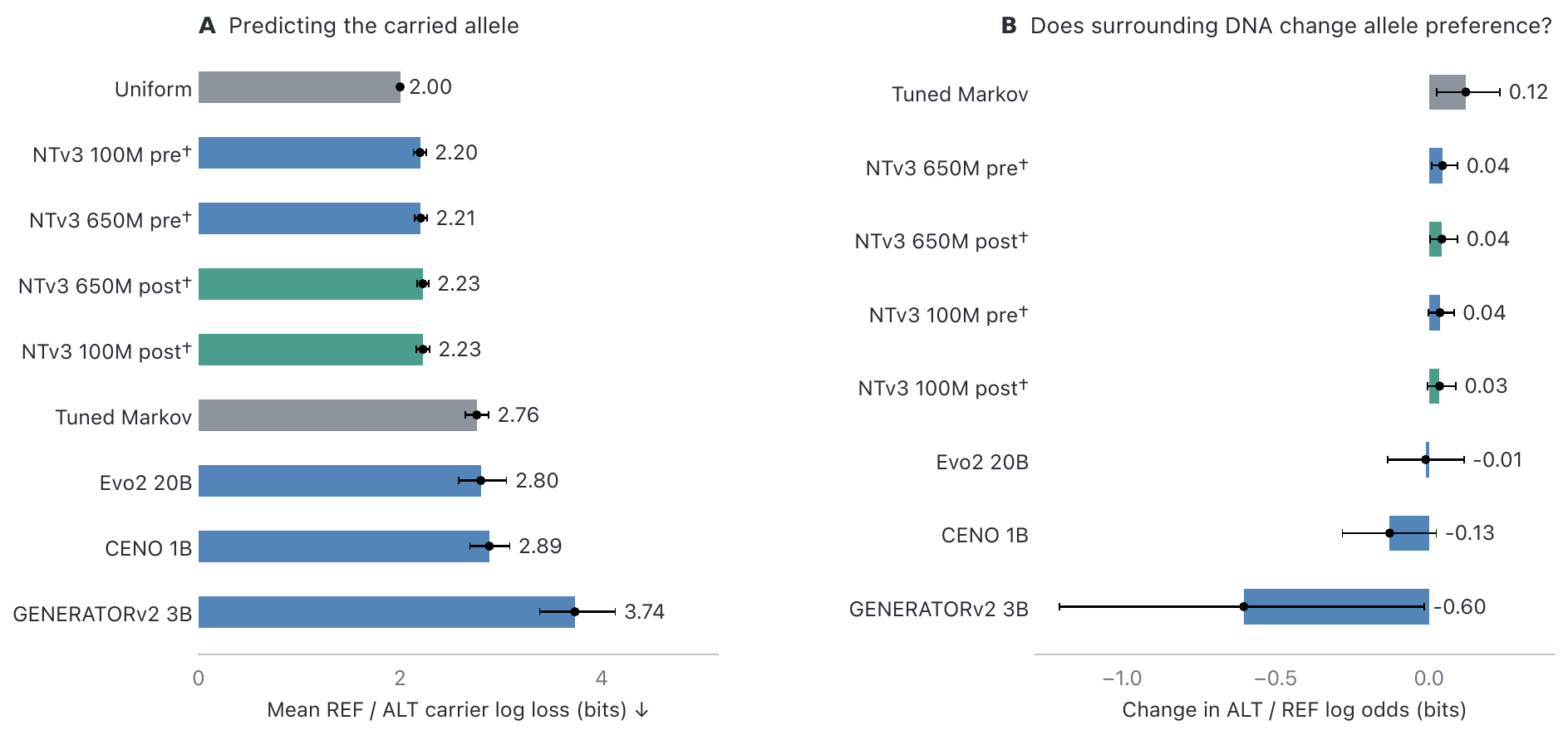}
  \capcaption{SequencePriors}{\caplead{SequencePriors}{Predicting alleles in natural haplotypes.}
  \emph{A:} base log loss at 512 SNP sites, with REF and ALT carriers equally
  weighted. Uniform prediction gives two bits.
  \emph{B:} change in ALT/REF log odds between carrier backgrounds for
  396 pairs with different surrounding DNA. Positive values indicate
  a shift towards ALT. NTv3 (\(\dagger\)) averages base probabilities from
  eight masked queries; causal models use uncalibrated probabilities.
  Whiskers show 95\% confidence intervals from resampling the 39 genomic blocks.}
  \label{fig:prior-population}
\end{figure}

Calibration reduces GENERATORv2's base loss from $3.735$ to $2.091$ bits. For NTv3 650M post, averaging eight masked queries instead of one reduces loss from $2.337$ to $2.225$ bits and raises
accuracy from 26.3\% to 36.3\%. All models shown still have higher
loss than uniform prediction, which gives two bits (Figure~\ref{fig:prior-population}A).

\paragraph{Allele preference and surrounding DNA.}
For each SNP, we score the same REF and ALT alleles in both carriers' surrounding DNA. We subtract the ALT/REF log odds in the REF background from those in the ALT background. Positive values mean a shift towards ALT.

Across the 396 pairs with different surrounding DNA, Evo2's preference changes little. NTv3 650M post shows a small shift towards ALT
($0.042$ bits, 95\% interval $0.004$--$0.093$;
Figure~\ref{fig:prior-population}B).

GENERATORv2's shift towards REF depends on where the SNP falls within a six-base token and on nearby indels. For NTv3, we mask corresponding positions in the two sequences.

\subsection{Sequence reconstruction and reporter effects}
\label{app:prior-functional}

We next evaluate whether models that recover DNA bases also predict how mutations affect expression. Table~\ref{tab:prior-assay-bridge} reports both measurements on the Siraj regulatory sequences, which are 189--200 bp long rather than the 1,200-bp genomic windows used for the main sequence-prior evaluation.

\paragraph{Reference-base recovery.}
At each site, we compare four otherwise identical sequences, with A, C, G or T at that position. 'Recovery' is the frequency with which the original base has
the highest full-sequence probability. Random guessing gives 25\%. We use 56,742 sites with all three alternative substitutions experimentally available.

\paragraph{Comparing sequence scores with RNA activity.}
For 170,292 mutations, we compare the change in log sequence probability with the measured increase or decrease in reporter RNA activity in HepG2 and K562, using Spearman correlation. This comparison also includes sites where one or two experimentally possible mutations are missing.

Log probabilities are averaged over the forward and reverse-complement strands. We calculate the metrics within each regulatory element, then average equally across 128 elements. We do not fit expression readouts for this comparison.

The results suggest that reconstructing DNA bases does not necessarily mean predicting how mutations affect expression.

\begin{table}[!htbp]
\centering\small
\setlength{\tabcolsep}{8pt}
\captable[84pt]{SequencePriors}
\begin{tabular}{lrr}
\toprule
Model & \shortstack{Reference-base\\recovery (\%)} & \shortstack{Correlation with RNA effects\\Spearman $\rho$} \\
\midrule
Evo2 20B & $55.2\;[52.5, 58.1]$ & $0.045\;[0.027, 0.065]$ \\
CENO 1B-1m & $62.4\;[58.2, 66.6]$ & $0.029\;[0.014, 0.044]$ \\
GENERATORv2 3B & $42.3\;[40.0, 44.7]$ & $0.034\;[0.018, 0.049]$ \\
\bottomrule
\end{tabular}
\caprender{SequencePriors}
\caption{\textbf{Base recovery and correlation with RNA effects.}
Brackets show 95\% confidence intervals from resampling regulatory
elements.}
\label{tab:prior-assay-bridge}
\end{table}

\FloatBarrier
\section{\capheading{FunctionalPrediction}{Variant-effect prediction and label efficiency}}
\label{app:simulate}

\paragraph[Training and evaluation.]{\phantomsection\label{app:model-comparisons}Training and evaluation.}
We use OpenSplice to evaluate how well genomic models predict mutation effects on RNA splicing (Figure~\ref{fig:simulate}A). We fit linear readouts to frozen model embeddings, training and testing on separate genes across five cross-validation splits. The 3--6-mer ridge baseline uses the same splits. We also evaluate the native splicing outputs of SpliceAI and AlphaGenome.

To test transfer to another assay, we train the readouts, CNN and ResNet on OpenSplice and evaluate them on COMPASS (Figure~\ref{fig:simulate}B). We exclude any overlapping sequences and remove the 23 shared genes from both datasets. The CNN and ResNet use 398 OpenSplice genes for training and another 45 for validation. Trivially, COMPASS measurements are not used for any kind of model selection, but only for evaluation.

\begin{figure}[!htbp]
  \centering
  \capfigure{FunctionalPrediction}
  \includegraphics[width=.78\linewidth]{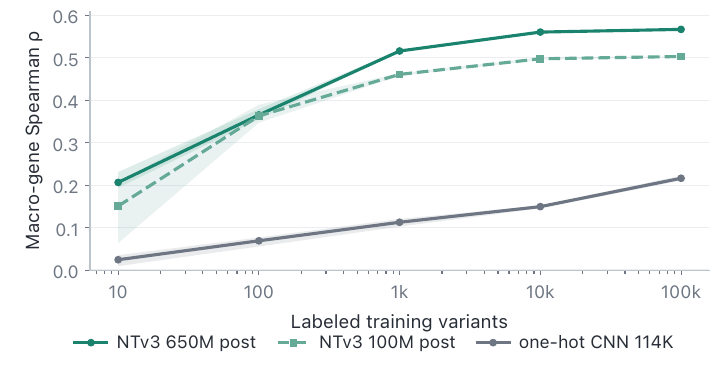}
  \capcaption{FunctionalPrediction}{\caplead{FunctionalPrediction}{OpenSplice prediction with fewer assay labels.}
  Larger training subsets include the smaller ones. Lines show the mean across three runs with different training samples and seeds while the shading shows their range. The training-set sizes count only
  OpenSplice examples. NTv3 post was previously trained on functional
  genomic data.}
  \label{fig:supp-label-efficiency}
\end{figure}

\paragraph{Learning from fewer examples.}
We vary the number of measured OpenSplice variants used to fit the NTv3 linear readouts and train the one-hot CNN. The models use the same training subsets and are evaluated on the same held-out genes. NTv3 readouts greatly outperform the CNN at every tested training-set size, showing strong gains from the use of foundation models (Figure~\ref{fig:supp-label-efficiency}).

\paragraph{Direction and magnitude of mutation effects.}
For mutations that have clearly measured effects ($|\Delta\mathrm{PSI}|>1.96\sigma$, where $\sigma$ is the reported measurement uncertainty), the NTv3 650M post readout correctly
predicts whether exon inclusion increases or decreases in 74.1\%
of cases, compared with 69.2\% for the one-hot ResNet.

Across all mutations, their root mean square errors (adjusted for measurement noise) are respectively 16.55 and 17.27 percentage points. We calculate each metric within genes and then average across genes. NTv3's mean calibration slope is
0.98 (95\% interval: 0.95--1.01), close to the ideal value of one,
indicating that predicted effects are on a very similar scale to the measured effects.

\FloatBarrier
\subsection{Changing focus from predicting effects to attempting an edit}
\label{app:editing-context}
Predicting mutation and edit effects is only part of DNA design: we also need to \textit{perform} the edit efficiently. We next test whether genomic models can predict how editing efficiency varies with the surrounding DNA.

\paragraph{Experimental setting.}
We use the K562 TRIP dataset from PRIDICT2 \citep{mathis2025pridict2}, which places a 640-bp reporter at 1,196 locations across 23 chromosomes. Editing efficiency is measured for prime editing, adenine editing (ABE8e-SpG), cytosine editing (BE4max) and Cas9 (SpCas9). Prime editing uses one fixed guide while the other three systems share another.

\paragraph[Prediction and evaluation.]{\phantomsection\label{app:editing-readouts}Prediction and evaluation.}
Models receive the same 8-kb reference-genome window around each integration site (without the inserted reporter). We fit a separate ridge readout to each model's frozen embeddings for each editing system. Training and testing use five cross-validation splits, keeping complete chromosomes together. Features are scaled using the training data, and the ridge penalty is fixed at $\alpha=1000$.

We additionally check whether the results depend on this penalty by choosing $\alpha\in\{1,10,\ldots,10^9\}$ through three-fold cross-validation within the training chromosomes. The chosen $\alpha$ minimizes mean squared prediction error.

We calculate Spearman correlation across the combined held-out predictions. Confidence intervals use 50,000 bootstrap resamples of whole chromosomes, without refitting the readouts. Comparisons between models are adjusted for multiple testing across all 312 comparisons.

\begin{figure}[!htbp]
  \centering
  \capfigure{FunctionalPrediction}
  \includegraphics[width=.80\linewidth]{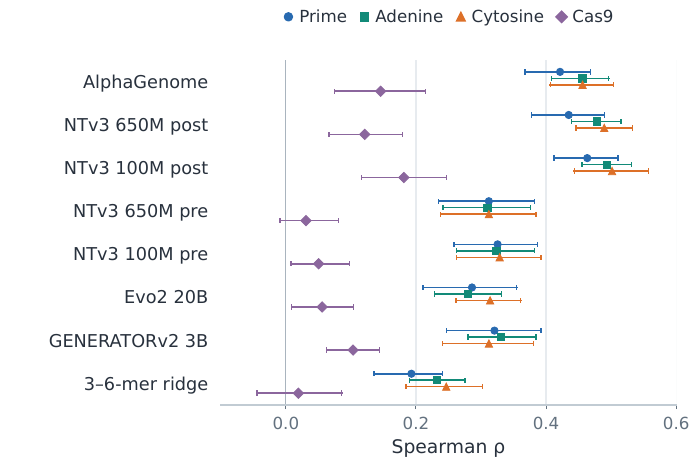}
  \capcaption{FunctionalPrediction}{\textbf{Predicting editing
  efficiency from surrounding DNA.}
  Spearman correlations for four editing systems in the K562 TRIP
  dataset. All models receive the same 8-kb genomic sequence;
  AlphaGenome receives additional N-padding on both sides to meet
  its 16-kb input requirement. Results are from five chromosome-held-out folds. Whiskers show pointwise 95\%
  confidence intervals from resampling chromosomes.}
  \label{fig:editing-context-models}
\end{figure}

\paragraph{Results.}
Functional training improves NTv3's predictions of prime- and base-editing efficiency at both model sizes (Figure~\ref{fig:editing-context-models}). This improvement remains when the ridge penalty is chosen by cross-validation. AlphaGenome and post-trained NTv3 give the strongest predictions, with no statistically significant differences between them. All models predict Cas9 efficiency substantially less accurately than prime- or base-editing efficiency.

\FloatBarrier
\section{\capheading{BiologicalContext}{Cell, tissue and personal-sequence effects}}
\label{app:context}

\subsection{Tissue expression contrasts and classification}
\label{app:gtex-tissue-contrast}

We evaluate expression predictions on 5,782 held-out genes across 51 GTEx tissues \citep{gtex2020atlas}, using $\log(1+\mathrm{TPM})$ expression.

\begin{figure}[htbp]
  \centering
  \capfigure{BiologicalContext}
  \includegraphics[width=\linewidth]{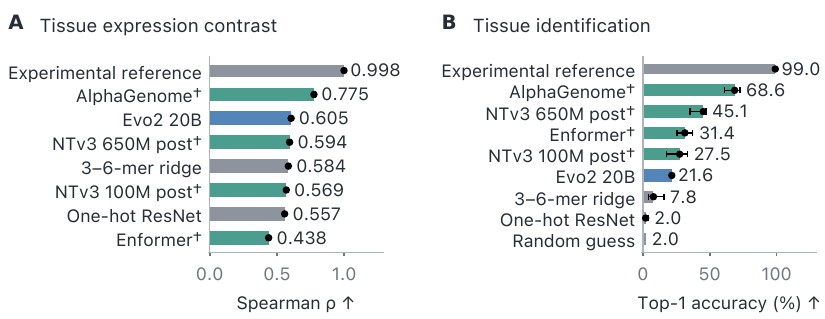}
  \capcaption{BiologicalContext}{\textbf{Predicting expression
  differences and identifying tissues.}
  A: correlation between predicted and measured expression
  differences within genes. Each tissue is paired with the tissue
  where measured expression differs most.
  B: tissue identification by matching predicted and measured
  expression profiles across genes (chance is $1/51$).
  Experimental references compare donor groups.
  $\dagger$ marks native outputs. Whiskers show available 95\%
  confidence intervals from resampling genes.}
  \label{fig:context-tissues}
\end{figure}

\paragraph{Expression differences.}
For each gene, we pair each tissue with the tissue where measured expression differs most. We then calculate the Spearman correlation between predicted and measured differences across all gene--tissue pairs (Figure~\ref{fig:context-tissues}A). Every model and the experimental reference use the same pairs.

\paragraph{Tissue identification.}
For each tissue, we compare its predicted expression profile across the test genes with the measured profiles of all 51 tissues. The tissue with the highest Spearman correlation is selected. We report the fraction of tissues correctly identified, sharing
credit for ties; chance is $1/51\approx 2\%$
(Figure~\ref{fig:context-tissues}B).

\paragraph{Readouts and evaluation.}
We compare native RNA predictions with linear readouts fitted to frozen model embeddings. For the fitted readouts, embeddings are averaged over the input sequence and projected to 512 features. Ridge regression uses 12,146 training genes, with the penalty chosen using 1,350 separate validation genes.

These readouts use 131,072-base windows centred on each gene's transcription start site. GENERATORv2 uses a shorter, centred 98,298-base window to fit its token limit. The evaluation holds out genes within the 51 tissues.

The experimental references compare expression measurements from two donor groups. Reported confidence intervals use 10,000 bootstrap resamples of whole genes, recomputing the rankings in each resample.

\subsection{Personal DNA at seen and unseen loci}
\label{app:personal-native}

We use MAGE data to test whether personal DNA improves gene-expression prediction for new people \citep{taylor2024mage}. We reconstruct both chromosome copies for 729 people and 557 chromosome-22 genes, using phased single-nucleotide variants.

We fit linear readouts to one-hot-encoded personal DNA. All fits also include donor and assay covariates. As a control, we repeat the fitting after shuffling whole genotypes among people from the same population and data split. Personal sequence gain is the difference in average within-gene correlation:
\begin{equation}
 \text{personal sequence gain}
 =\rho_{\G}(\widehat y_{\mathrm{personal}},y)
 -\rho_{\G}(\widehat y_{\mathrm{shuffled}},y).
 \label{eq:personal}
\end{equation}
Both comparisons hold out people. The second also holds out gene regions, keeping overlapping 8-kb windows together when splitting the data.

\begin{figure}[t]
  \centering
  \setlength{\belowcaptionskip}{12pt}
  \capfigure{BiologicalContext}
  \includegraphics[width=.91\linewidth]{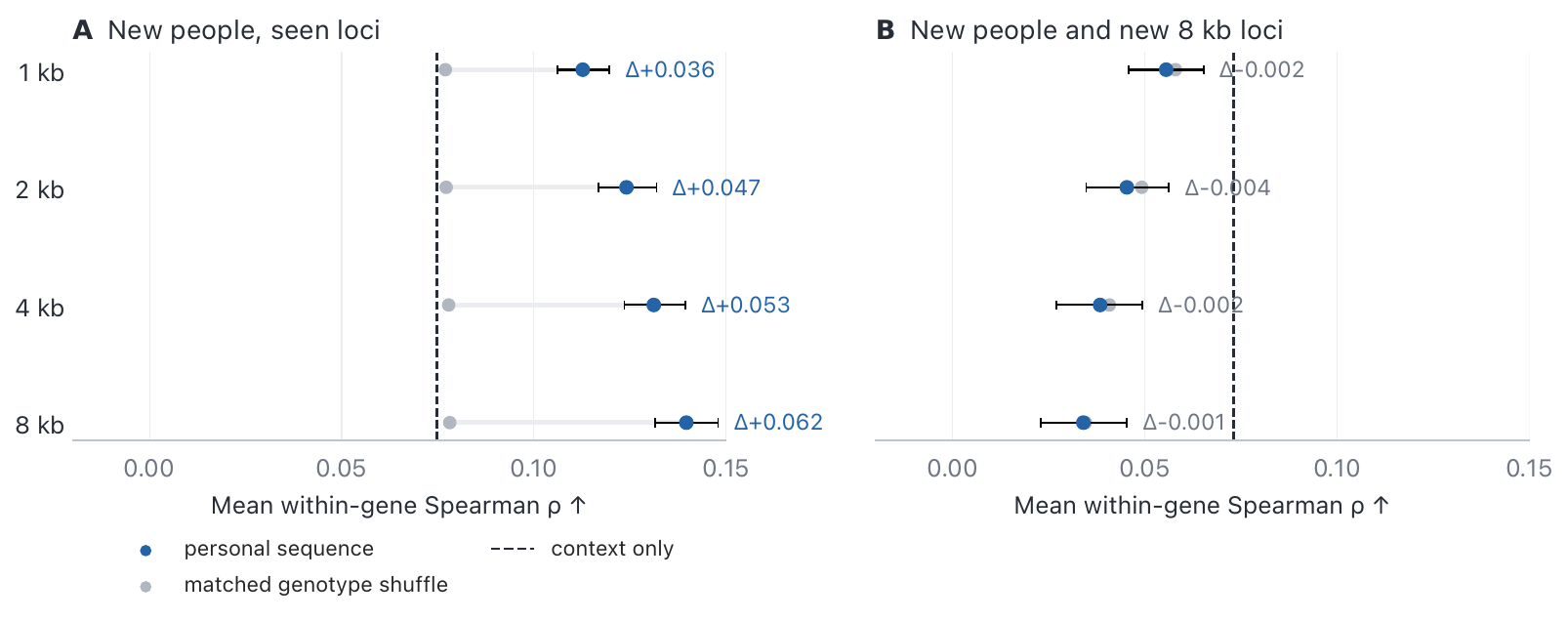}
  \capcaption{BiologicalContext}{\caplead{BiologicalContext}{Personal DNA helps at familiar genes.}
  Expression-prediction correlations for new
  people at genes seen during training (left) or unseen gene regions
  (right). $\Delta$ gives the gain over shuffled genotypes. Intervals are obtained by resampling genes.}
  \label{fig:original-personal}
\end{figure}

At genes seen during training, using the correct personal DNA improves prediction: the gain rises from $0.036$ with 1-kb windows to $0.062$ with 8-kb windows (Figure~\ref{fig:original-personal}, left). When the gene region is also unseen, all confidence intervals however include zero, with an 8-kb gain of practically zero (right). Personal DNA thus helps when the readout has learned from the same genes in other people, but shows no clear benefit at new gene regions.

\FloatBarrier
\section{\capheading{GeneticInteractions}{Interaction prediction}}
\label{app:genetic-interactions}
\subsection{Predicting COMPASS mutation effects.} \label{app:compass-predictions}

We use the linear-readout method described in Appendix~\ref{app:model-comparisons}, training on OpenSplice substitutions. Here, we exclude exact sequence overlaps but retain genes shared between the datasets. The interaction analysis includes 15,035 single-mutant and 48,054 double-mutant measurements in 110 genes (Figure~\ref{fig:compose}B).

Before calculating interactions, we subtract each readout's prediction for an unchanged sequence, so that an unchanged sequence has zero predicted effect. Figure confidence intervals use 10,000 bootstrap samples of whole genes.

\subsection{How much do predicted interactions contribute?}
\label{app:interaction-ablation}

For COMPASS, we replace each predicted double-mutant effect with the sum of its two single-mutant effects:
\[
 \widehat\Delta^{\mathrm{abl}}(\delta+\epsilon)
=\widehat\Delta(\delta)+\widehat\Delta(\epsilon).
\]

For Siraj, we replace a mutation's predicted effects on its two backgrounds with their mean:
\[
 \widehat\Delta_A^{\mathrm{abl}}
=\widehat\Delta_R^{\mathrm{abl}} =(\widehat\Delta_A+\widehat\Delta_R)/2.
\]
We use 68,026 paired mutations in 114 regulatory elements, averaging measured effects across HepG2 and K562.

We calculate Spearman correlations within genes or regulatory elements and then average them. Confidence intervals use 20,000 paired bootstrap samples of whole genes or elements, with a Bonferroni correction for 22 correlation and RMSE comparisons.

\begin{table}[ht]
  \centering
  \small
  \setlength{\belowcaptionskip}{12pt}
  \setlength{\tabcolsep}{5.5pt}
  \renewcommand{\arraystretch}{1.12}
  \newlength{\interactionScoreWidth}
  \settowidth{\interactionScoreWidth}{Without interaction}
  \captable{GeneticInteractions}
  \begin{tabular}{@{}lccrl@{}}
    \toprule
    Model &
    \makebox[\interactionScoreWidth][c]{Original} &
    \makebox[\interactionScoreWidth][c]{Without interaction} &
    \multicolumn{1}{c}{Difference} &
    \multicolumn{1}{c@{}}{95\% interval} \\[-2pt]
    & $\rho$ & $\rho$ &
    \multicolumn{1}{c}{$\Delta\rho$} & \\
    \midrule
    \multicolumn{5}{@{}l}{\emph{COMPASS: double-mutant effects}} \\
    \addlinespace[2pt]
    AlphaGenome   & $0.471$ & $0.462$ & $+0.0098\phantom{0}$ & $[-0.0059,\phantom{-}0.0245]$ \\
    SpliceAI      & $0.555$ & $0.559$ & $-0.0039\phantom{0}$ & $[-0.0181,\phantom{-}0.0044]$ \\
    NTv3 100M post & $0.423$ & $0.418$ & $+0.0050\phantom{0}$ & $[-0.0109,\phantom{-}0.0219]$ \\
    NTv3 650M post & $0.500$ & $0.493$ & $+0.0070\phantom{0}$ & $[\phantom{-}0.0014,\phantom{-}0.0144]$ \\
    GENERATORv2   & $0.257$ & $0.243$ & $+0.0134\phantom{0}$ & $[-0.0066,\phantom{-}0.0307]$ \\
    \addlinespace[5pt]
    \multicolumn{5}{@{}l}{\emph{Siraj: mutation effects on paired backgrounds}} \\
    \addlinespace[2pt]
    AlphaGenome   & $0.528$ & $0.521$ & $+0.0070\phantom{0}$ & $[\phantom{-}0.0014,\phantom{-}0.0120]$ \\
    Enformer      & $0.450$ & $0.451$ & $-0.0006\phantom{0}$ & $[-0.0064,\phantom{-}0.0046]$ \\
    NTv3 100M post & $0.417$ & $0.417$ & $-0.00003$ & $[-0.0057,\phantom{-}0.0051]$ \\
    NTv3 650M post & $0.487$ & $0.481$ & $+0.0055\phantom{0}$ & $[\phantom{-}0.0009,\phantom{-}0.0099]$ \\
    GENERATORv2   & $0.208$ & $0.211$ & $-0.0039\phantom{0}$ & $[-0.0077,-0.0001]$ \\
    Evo2 20B      & $0.206$ & $0.209$ & $-0.0027\phantom{0}$ & $[-0.0068,\phantom{-}0.0013]$ \\
    \bottomrule
  \end{tabular}
  \caprender{GeneticInteractions}
  \caption{\textbf{Effect prediction with and without interactions.}
  Overall effect accuracy changes little when predicted interactions are removed. Spearman correlations are averaged across genes for COMPASS and
regulatory elements for Siraj. Positive differences indicate better
accuracy with interactions included; brackets show 95\% confidence
intervals adjusted for multiple comparisons.}
  \label{tab:interaction-ablation}
\end{table}

AlphaGenome predicts Siraj interactions with correlation $0.260$, yet removing interactions changes its overall effect correlation by only $0.007$ (Table~\ref{tab:interaction-ablation}). A high correlation for overall effects is therefore not enough to show that a model predicts interactions well.

\subsection{Fine-tuning for interaction prediction}
\label{app:conditional-development}

We test whether fine-tuning NTv3 100M post improves interaction prediction compared with a readout fitted to its frozen embeddings. Both approaches use the same Siraj measurements, with separate training, validation and test loci. We update the final decoder block using mutations measured on both backgrounds at 68 training loci and evaluate on 23 test loci. Results average three training runs, and confidence intervals resample test loci.

Fine-tuning improves effect correlation by $0.065$
(95\% interval $0.005$--$0.125$), but does not clearly improve
interaction correlation (change $-0.016$, 95\% interval
$-0.040$--$0.007$).

\FloatBarrier
\section{\capheading{GenomicDesign}{Target matching and selective expression}}
\label{app:sequence-choice}

\subsection{Matching a target response}
\label{app:response-matching}

In the Siraj assay, we start from a regulatory element's reference sequence. The reference edit $\delta_0$ introduces an alternative allele. We ask whether a single-base edit at any other position in the reference sequence can produce a similar change in RNA activity. We also test the reverse direction. Notably, Siraj has experimentally measured the entire design space, thus enabling this task.

We retain 331 targets across 110 elements for which the measured reference-edit effect is at least twice its standard error in magnitude. Each model ranks candidate edits by how closely their predicted effects match its own prediction for $\delta_0$.

For the first selected edit, we report the percentile $100(r-1)/(N-1)$, where $r$ ranks its measured distance from the target among $N$ eligible edits. Lower percentiles are hence closer matches.

\paragraph{Success within a testing budget.}
For a budget of $B$ tests, $\operatorname{TopB}_m$ contains the first $B$ edits selected by model $m$. A target is matched if at least one selected edit differs from the measured reference effect by no more than one combined standard error:
\begin{equation}
  S_B=\mathbf{1}\!\left\{
  \min_{\eta\in\operatorname{TopB}_m}
  \frac{\left|\Delta(\eta\mid x,c)-\Delta(\delta_0\mid x,c)\right|}
  {\sqrt{\mathrm{SE}(\eta)^2+\mathrm{SE}(\delta_0)^2}}
  \leq 1
  \right\}.
  \label{eq:repair}
\end{equation}
Here, $\Delta$ is the measured change in RNA activity and $\mathrm{SE}$ its standard error. Both percentile ranks and success rates are averaged over targets within each element, then equally across elements.

\subsection{Selecting mutations for one allelic background}
\label{app:final-siraj}

We also test whether AlphaGenome can choose mutations that change expression on one allelic background while leaving the others largely unchanged. The main Siraj panel contains 114 loci with two backgrounds. Two smaller panels contain 11 loci with two backgrounds and 13 loci with four backgrounds.

\paragraph[Choosing mutations.]{\phantomsection\label{app:selective-candidates}Choosing mutations.}
Mutation candidates are substitutions within ten bases of the natural variant positions (excluding the natural variant position). For each cell, target background $t$ and desired direction $d$ ($+1$ for an increase, $-1$ for a decrease), we select the mutation with the highest predicted utility:
\begin{equation}
 U(\epsilon)=d\,\Delta_t(\epsilon)
 -\frac{1}{B-1}\sum_{b\ne t}|\Delta_b(\epsilon)|,
 \qquad U(\text{no edit})=0,
 \label{eq:selective-utility}
\end{equation}
where $B$ is the number of backgrounds. We choose no edit if all predicted utilities are non-positive. Results are averaged within loci and then equally across loci.

\paragraph{Model predictions.}
AlphaGenome receives each experimental haplotype in its genomic context, using 131,072-base inputs. We sum predicted accessibility changes over 501 bases centered on the mutation, normalize each track by its supplied mean, and average the cell-matched ATAC and DNase predictions.

\paragraph{Measured RNA effects.}
We evaluate repeat measurements 1--2 and 3--5 separately. After summing barcode counts for each sequence, we calculate the $\log_2$ RNA-to-plasmid ratio with a pseudocount of $0.5$ and average across repeats. Within each cell and background, we subtract the mean activity of sequences between the first and third activity quartiles. This reference includes both mutated and unmutated sequences.

\paragraph[Matching the target response.]{\phantomsection\label{app:selective-control}Matching the target response.}
To check whether selectivity simply comes from choosing fewer or weaker edits, we compare with a control that ignores effects on the other backgrounds. It chooses the same number of edits within each locus and cell, although the target backgrounds and directions may differ.

The control combines five selection rules covering weaker and stronger predicted target effects. Their proportions are fitted using measured target responses at other loci, aiming to match the selection method's mean target response. For the 114-locus panel, we use five folds, keeping evaluation loci out of this fitting. Controls for the smaller panels are fitted using the 114-locus panel.

Confidence intervals use 100,000 bootstrap samples of whole loci, refitting the control in each sample. We adjust the confidence intervals for our comparisons using Bonferroni correction.

\begin{figure}[!htbp]
 \centering
 \capfigure{GenomicDesign}
 \includegraphics[width=\linewidth]{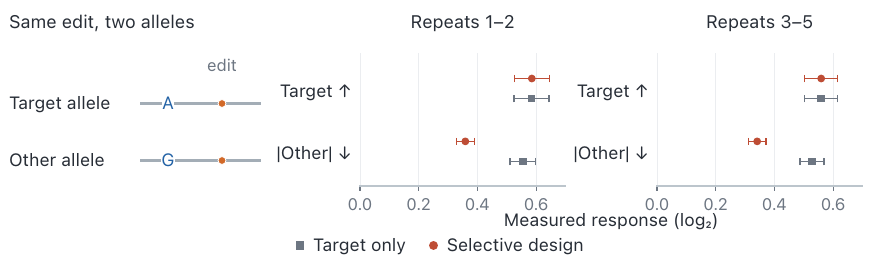}
 \caprender{GenomicDesign}
 \caption{\textbf{Reducing effects on the other allelic background.}
 Within the same cell type, we compare the target response and
the absolute response on the other allelic background for
mutations selected by AlphaGenome at 114 Siraj loci. The control matches the number of edits and mean target response.
 Results are shown separately for repeats 1--2 and 3--5 while
 the same mutation is evaluated on both backgrounds.
 Whiskers show unadjusted 95\% confidence intervals.}
 \label{fig:allele-selective}
\end{figure}

\paragraph{Results.}
On the 114-locus panel, selective choices reduce the absolute
response on the other background by about $35\%$ in both repeat
groups (Figure~\ref{fig:allele-selective}). Mean target responses
remain similar: adjusted 95\% intervals for their differences
lie entirely within $\pm0.05\,\log_2$ units.
Utility gains are $0.196$ (adjusted 95\% interval
$0.139$--$0.258$) and $0.186$ ($0.131$--$0.243$).

\subsection{Selecting mutations for one cell type}
\label{app:cell-selective}

For the cell-selective design task in Figure~\ref{fig:control}C, we use the Siraj HepG2/K562 assay \citep{siraj2026functional} to choose mutations that increase expression in one cell type while limiting changes in the other. The analysis includes 167,958 substitutions across 280 parental backgrounds and 127 loci; one locus lacks the required count data. Both cells serve as targets, and all measured substitutions are eligible, including natural-allele substitutions and reversions.

For each repeat, we compare the mutation's RNA-to-plasmid ratio with that of its unmutated parent, then average across repeats:
\begin{equation}
 z_{s,c,r}=\log_2\frac{\mathrm{RNA}_{s,c,r}+0.5}
                         {\mathrm{DNA}_{s,r}+0.5},
 \qquad
 \Delta_c(s)=\frac{1}{5}\sum_{r=1}^{5}
       (z_{s,c,r}-z_{m0,c,r}),
\end{equation}
where $m0$ denotes the unmutated parent. We also calculate effects separately for repeats 1--2 and 3--5 for the experimental reference below.

\paragraph[Predicting effects.]{\phantomsection\label{app:cell-selective-fitting}Predicting effects.}
We compare AlphaGenome and Enformer native accessibility outputs, NTv3 100M/650M post readouts, a 3--6-mer ridge baseline and a one-hot ResNet. AlphaGenome and Enformer predictions are summed over 501-bp and 1,024-bp windows, respectively. We convert these accessibility predictions to RNA-effect units by fitting a nonnegative scale factor for each cell. Five folds keep complete loci together, and calibration uses only loci outside the evaluation fold. Training loci receive equal weight.

For NTv3 and the baselines, we combine the predicted mean effect across cells with the predicted HepG2-minus-K562 difference: HepG2 receives the mean plus half the difference, and K562 the mean minus half the difference. These readouts were trained on the published RNA-effect measurements and are evaluated here against effects relative to the unmutated parent, without further calibration.

\paragraph{Choosing and evaluating edits.}
For each parent and target cell, we choose the substitution maximizing $\widehat\Delta_{\mathrm{target}} -\lambda|\widehat\Delta_{\mathrm{other}}|$, using $\lambda\in\{0,0.25,0.5,1,2,4\}$. We choose no edit if no mutation has positive predicted utility. The main comparison uses $\lambda=1$.

Measured utility is the target-cell effect minus the absolute other-cell effect. We average these values across backgrounds and target cells within each locus, then equally across loci. Choosing no edit contributes zero.

The control makes an edit for exactly the same parent and target-cell combinations as the selective method. It mixes six target-only selection rules covering weaker and stronger predicted effects, matching the selective method's average predicted target effect within each combination. Other-cell predictions do not rank the control's candidates, and measured responses are not used to fit the mixture. The matching therefore applies to predictions; measured RNA effects can still differ.

We use 50,000 bootstrap samples of whole loci, keeping fitted models, calibration factors and selected edits fixed. Confidence intervals for the 24 comparisons across the six models are adjusted using Bonferroni correction.

AlphaGenome, Enformer and both post-trained NTv3 readouts achieve positive utility. AlphaGenome and both NTv3 readouts also improve over their matched controls. Enformer's gain over its control is uncertain:
$0.055$ (adjusted 95\% interval $-0.011$--$0.120$).

\paragraph{Experimental reference.}
We choose edits using measurements from repeats 1--2 and evaluate them using repeats 3--5, then reverse the groups and average the results. Both directions use the same candidates, penalties, no-edit option and control procedure.

The reference reaches mean utility $0.962$
(95\% locus-bootstrap interval $0.829$--$1.092$),
with gains of $0.371$ over target-only selection and $0.292$ over the matched control. On average, edits chosen using one group of measurements remain selective in the other.

\subsection{Selecting enhancers for multi-cell expression programs}
\label{app:multicell-programming}

The multi-cell task in Figure~\ref{fig:frontier-programming} uses Atlas reporter measurements \citep{castillohair2025atlas} (\href{https://zenodo.org/records/17410822}{Zenodo record 17410822}). We evaluate sixteen expression patterns: eight with two cell types ON and eight with three ON (with the remaining cells OFF). Each pattern uses the same 2,411 distinct 145-nt candidate sequences, regardless of their original design target.

We use twelve cell types with two RNA measurements each. For each cell and its repeat, the activity is the log RNA-to-DNA ratio relative to the median of the 35 negative controls:
\[
y_{icr}=\log_2\frac{R_{icr}+0.5}{D_{icr}+0.5} -\operatorname{median}_{j\in\mathcal N}
   \log_2\frac{R_{jcr}+0.5}{D_{jcr}+0.5}.
\]
Counts are UMI-deduplicated, and we retain the authors' DNA coverage filter of more than 20 reads in each preparation. The two RNA measurements stem from two separate experimental transfections.

\paragraph{Model predictions.}
We add the assay's cloning junctions and pad with Ns to 16,384 bases for AlphaGenome, 196,608 for Enformer and 1,024 for NTv3 post. We average each model's native human outputs over the insert and apply a $\log(1+x)$ transform. Using the mean of the two RNA measurements for fitting, linear ridge readouts then predict activity in all twelve cells.

Sequences sharing a genomic background or high sequence similarity are grouped into families. Excluding families shared with the candidates leaves 5,246 sequences for fitting. We train on 4,223 and choose the ridge penalty using 1,023 validation sequences, keeping families together. We then refit on all 5,246 sequences. The readout fitted to AlphaGenome's frozen embeddings uses the same data and fitting procedure.

\paragraph{Selection and evaluation.}
For each expression pattern, models select the ten sequences with the highest predicted utility (Equation~\ref{eq:multicell-utility}). We evaluate the same choices using each RNA measurement, averaging over the ten sequences and then equally across the sixteen patterns.

\begin{figure}[!htbp]
 \centering
 \capfigure{GenomicDesign}
 \includegraphics[width=.92\linewidth]{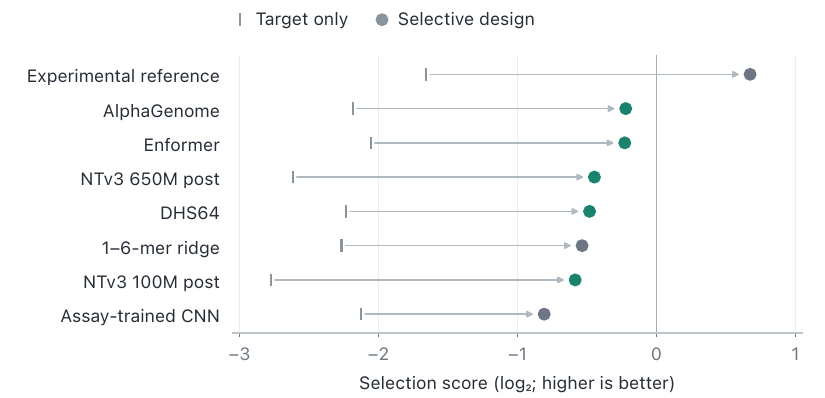}
 \capcaption{GenomicDesign}{\textbf{Considering unwanted activity
 improves sequence selection.}
 Arrows run from selection based only on the weakest ON-cell
 prediction (ticks) to selection that also penalizes the
 strongest OFF-cell prediction (dots). Both are scored using
 Equation~\ref{eq:multicell-utility}, averaged over ten sequences,
 sixteen expression patterns and both RNA measurements.
 The experimental reference selects using one measurement and
 evaluates using the other.}
 \label{fig:atlas-selector-control}
\end{figure}

We compare these choices with selection based only on the weakest ON-cell prediction (Figure~\ref{fig:atlas-selector-control}), random selection from all candidates, and random selection among sequences which the Atlas authors originally designed for the same ON/OFF pattern. Figure~\ref{fig:frontier-programming}A shows gains over this last difficult control, using one point for each RNA measurement.

The experimental reference selects sequences using one RNA measurement and evaluates them using the other, then swaps the measurements. Figure~\ref{fig:frontier-programming}B shows one example: AlphaGenome's ten sequence choices for HepG2/K562, ordered by predicted utility, with activities averaged across both measurements.

\clearpage
\section{Quantification of uncertainty}
\label{app:statistical-comparisons}
\label{app:bend2-reference}
\label{app:benchmark-measures}
\definecolor{benchInk}{HTML}{26323D}
\definecolor{benchRail}{HTML}{FFFFFF}
\definecolor{benchRailHead}{HTML}{FFFFFF}
\definecolor{benchPrior}{HTML}{A18338}
\definecolor{benchSimulate}{HTML}{477FA5}
\definecolor{benchContext}{HTML}{438879}
\definecolor{benchCompose}{HTML}{8266A2}
\definecolor{benchControl}{HTML}{B56E58}
\definecolor{unInk}{HTML}{26323D}
\newcommand{\unNative}[1]{\hspace{6pt}\textcolor{black!65}{#1}}
\newcommand{\unCell}[2]{\cellcolor{#1}\mbox{#2}\hspace{3.5pt}\mbox{}}
\newcommand{\unEmpty}[1]{\cellcolor{#1}\textcolor{unInk!27}{---}\hspace{3.5pt}\mbox{}}
\newcommand{\unHeader}[3]{%
  \cellcolor{bench#1!6}%
  {\fontsize{8.1}{9.4}\selectfont
  \makebox[\linewidth][c]{\shortstack{\rule{0pt}{9pt}#2\\[2pt]%
  {\fontsize{7.3}{8.5}\selectfont\color{unInk!80}#3\rule[-2pt]{0pt}{2pt}}}}}}
\newenvironment{uncertaintytable}{%
  \begingroup\fontfamily{phv}\fontsize{8.5}{10}\selectfont\color{unInk}%
  \setlength{\tabcolsep}{0pt}\renewcommand{\arraystretch}{1}%
}{\endgroup}

\subsection{Confidence intervals}
\label{app:confidence-intervals}
\label{app:final-inference}
The intervals and point estimates below correspond to Table~\ref{tab:bend2-reference}.

\begin{table}[!ht]
\centering
\begin{uncertaintytable}
\begin{tabular}{@{}>{\raggedright\arraybackslash}p{\dimexpr.26\linewidth-6pt\relax}@{\hspace{6pt}}>{\raggedleft\arraybackslash}p{\dimexpr.185\linewidth-4.5pt\relax}@{\hspace{6pt}}>{\raggedleft\arraybackslash}p{\dimexpr.185\linewidth-4.5pt\relax}@{\hspace{6pt}}>{\raggedleft\arraybackslash}p{\dimexpr.185\linewidth-4.5pt\relax}@{\hspace{6pt}}>{\raggedleft\arraybackslash}p{\dimexpr.185\linewidth-4.5pt\relax}@{}}
\textcolor{unInk!80}{Model} & \unHeader{Prior}{Prior\\loss}{bits $\downarrow$} & \unHeader{Simulate}{Splicing\\effects}{$\rho\,\uparrow$} & \unHeader{Context}{Cell\\contrast}{$\rho\,\uparrow$} & \unHeader{Context}{Tissue\\ID}{\% $\uparrow$}\\[3pt]
\cellcolor{white}Base composition & \unCell{benchPrior!5.00}{[1.953, 1.997]} & \unEmpty{benchSimulate!5.00} & \unEmpty{benchContext!5.00} & \unEmpty{benchContext!5.00}\\[0pt]
\cellcolor{white}Evo2 20B & \unCell{benchPrior!31.47}{[1.250, 1.538]} & \unCell{benchSimulate!6.86}{[0.352, 0.381]} & \unCell{benchContext!5.39}{[0.022, 0.060]} & \unCell{benchContext!6.98}{21.6}\\[0pt]
\cellcolor{white}CENO 1B & \unCell{benchPrior!7.99}{[1.385, 1.736]} & \unCell{benchSimulate!5.56}{[0.282, 0.309]} & \unCell{benchContext!5.16}{[0.018, 0.053]} & \unCell{benchContext!5.70}{[5.9, 19.6]}\\[0pt]
\cellcolor{white}GENERATORv2 & \unCell{benchPrior!5.00}{[1.652, 1.775]} & \unCell{benchSimulate!7.70}{[0.374, 0.402]} & \unCell{benchContext!5.78}{[0.050, 0.083]} & \unCell{benchContext!6.45}{[9.8, 21.6]}\\[0pt]
\cellcolor{white}DNABERT-2 & \unEmpty{benchPrior!5.00} & \unCell{benchSimulate!5.16}{[0.188, 0.210]} & \unCell{benchContext!6.28}{[0.066, 0.087]} & \unCell{benchContext!5.06}{[2.0, 9.8]}\\[0pt]
\addlinespace[2.2pt]
\cellcolor{white}NTv3 100M pre & \unCell{benchPrior!11.06}{[1.454, 1.620]} & \unCell{benchSimulate!6.21}{[0.335, 0.363]} & \unCell{benchContext!6.28}{[0.063, 0.089]} & \unCell{benchContext!5.28}{[5.9, 15.7]}\\[0pt]
\cellcolor{white}NTv3 100M post & \unCell{benchPrior!15.15}{[1.393, 1.592]} & \unCell{benchSimulate!12.53}{[0.488, 0.516]} & \unCell{benchContext!11.81}{[0.215, 0.273]} & \unCell{benchContext!16.34}{[35.3, 56.9]}\\[0pt]
\cellcolor{white}\unNative{Native outputs} & \unEmpty{benchPrior!5.00} & \unCell{benchSimulate!5.00}{[0.151, 0.185]} & \unCell{benchContext!8.06}{[0.168, 0.237]} & \unCell{benchContext!7.76}{[17.6, 33.3]}\\[0pt]
\addlinespace[1.8pt]
\cellcolor{white}NTv3 650M pre & \unCell{benchPrior!22.22}{[1.352, 1.539]} & \unCell{benchSimulate!5.03}{[0.174, 0.199]} & \unCell{benchContext!5.00}{[-0.001, 0.023]} & \unCell{benchContext!5.28}{[3.9, 11.8]}\\[0pt]
\cellcolor{white}NTv3 650M post & \unCell{benchPrior!38.00}{[1.275, 1.475]} & \unCell{benchSimulate!17.14}{[0.550, 0.578]} & \unCell{benchContext!14.85}{[0.229, 0.283]} & \unCell{benchContext!22.97}{[45.1, 62.7]}\\[0pt]
\cellcolor{white}\unNative{Native outputs} & \unEmpty{benchPrior!5.00} & \unCell{benchSimulate!9.57}{[0.445, 0.477]} & \unCell{benchContext!10.57}{[0.207, 0.271]} & \unCell{benchContext!12.38}{[35.3, 47.1]}\\[0pt]
\addlinespace[2.2pt]
\cellcolor{white}AlphaGenome & \unEmpty{benchPrior!5.00} & \unCell{benchSimulate!31.23}{[0.634, 0.665]} & \unCell{benchContext!38.00}{[0.371, 0.433]} & \unCell{benchContext!30.93}{[51.0, 68.6]}\\[0pt]
\cellcolor{white}\unNative{Native outputs} & \unEmpty{benchPrior!5.00} & \unCell{benchSimulate!24.51}{[0.605, 0.637]} & \unCell{benchContext!32.27}{[0.366, 0.431]} & \unCell{benchContext!38.00}{[60.8, 72.5]}\\[0pt]
\cellcolor{white}Enformer & \unEmpty{benchPrior!5.00} & \unEmpty{benchSimulate!5.00} & \unCell{benchContext!20.69}{[0.242, 0.321]} & \unCell{benchContext!9.17}{[25.5, 37.3]}\\[0pt]
\addlinespace[2.2pt]
\cellcolor{white}$k$-mer ridge & \unEmpty{benchPrior!5.00} & \unCell{benchSimulate!5.65}{[0.307, 0.329]} & \unCell{benchContext!9.00}{[0.194, 0.228]} & \unCell{benchContext!5.28}{[3.9, 15.7]}\\[0pt]
\cellcolor{white}Task-trained ResNet & \unEmpty{benchPrior!5.00} & \unCell{benchSimulate!11.09}{[0.477, 0.498]} & \unCell{benchContext!5.03}{[0.011, 0.029]} & \unCell{benchContext!5.00}{[2.0, 3.9]}\\[0pt]
\cellcolor{white}SpliceAI & \unEmpty{benchPrior!5.00} & \unCell{benchSimulate!38.00}{[0.659, 0.689]} & \unEmpty{benchContext!5.00} & \unEmpty{benchContext!5.00}\\[0pt]
\addlinespace[3.0pt]
\cellcolor{white}\textcolor{black!65}{Experimental reference} & \unEmpty{black!3} & \unCell{black!3}{[0.792, 0.825]} & \unCell{black!3}{[0.567, 0.620]} & \unCell{black!3}{99.0}\\[0pt]
\end{tabular}\end{uncertaintytable}

\par\vspace{7pt}
\begin{uncertaintytable}
\begin{tabular}{@{}>{\raggedright\arraybackslash}p{\dimexpr.26\linewidth-6pt\relax}@{\hspace{6pt}}>{\raggedleft\arraybackslash}p{\dimexpr.185\linewidth-4.5pt\relax}@{\hspace{6pt}}>{\raggedleft\arraybackslash}p{\dimexpr.185\linewidth-4.5pt\relax}@{\hspace{6pt}}>{\raggedleft\arraybackslash}p{\dimexpr.185\linewidth-4.5pt\relax}@{\hspace{6pt}}>{\raggedleft\arraybackslash}p{\dimexpr.185\linewidth-4.5pt\relax}@{}}
\textcolor{unInk!80}{Model} & \unHeader{Compose}{Background\\interaction}{$\rho\,\uparrow$} & \unHeader{Control}{Single-edit\\rank}{percentile $\downarrow$} & \unHeader{Control}{Cell\\selectivity}{$U\,\uparrow$} & \unHeader{Control}{Multi-cell\\gain}{log$_2$ $\uparrow$}\\[3pt]
\cellcolor{white}Base composition & \unEmpty{benchCompose!5.00} & \unEmpty{benchControl!5.00} & \unEmpty{benchControl!5.00} & \unEmpty{benchControl!5.00}\\[0pt]
\cellcolor{white}Evo2 20B & \unCell{benchCompose!6.92}{[0.034, 0.066]} & \unCell{benchControl!7.27}{[39.5, 47.5]} & \unCell{benchControl!5.00}{[-0.418, -0.286]} & \unCell{benchControl!7.63}{[0.233, 0.317]}\\[0pt]
\cellcolor{white}CENO 1B & \unCell{benchCompose!5.16}{[-0.007, 0.027]} & \unCell{benchControl!5.03}{[44.9, 52.7]} & \unCell{benchControl!5.03}{[-0.392, -0.248]} & \unCell{benchControl!5.00}{[-0.584, -0.579]}\\[0pt]
\cellcolor{white}GENERATORv2 & \unCell{benchCompose!6.29}{[0.030, 0.062]} & \unCell{benchControl!6.57}{[40.9, 48.5]} & \unCell{benchControl!5.32}{[-0.326, -0.190]} & \unCell{benchControl!7.03}{[0.226, 0.243]}\\[0pt]
\cellcolor{white}DNABERT-2 & \unCell{benchCompose!5.39}{[0.007, 0.032]} & \unCell{benchControl!5.13}{[45.0, 52.2]} & \unCell{benchControl!5.13}{[-0.377, -0.230]} & \unCell{benchControl!6.79}{[0.224, 0.242]}\\[0pt]
\addlinespace[2.2pt]
\cellcolor{white}NTv3 100M pre & \unCell{benchCompose!5.16}{[0.000, 0.021]} & \unCell{benchControl!5.32}{[43.8, 50.8]} & \unCell{benchControl!6.02}{[-0.265, -0.151]} & \unCell{benchControl!8.54}{[0.332, 0.347]}\\[0pt]
\cellcolor{white}NTv3 100M post & \unCell{benchCompose!9.05}{[0.110, 0.168]} & \unCell{benchControl!14.14}{[25.9, 34.8]} & \unCell{benchControl!11.38}{[0.124, 0.299]} & \unCell{benchControl!6.18}{[0.106, 0.182]}\\[0pt]
\cellcolor{white}\unNative{Native outputs} & \unCell{benchCompose!10.40}{[0.123, 0.182]} & \unCell{benchControl!11.22}{[27.9, 36.2]} & \unCell{benchControl!10.35}{[0.101, 0.302]} & \unCell{benchControl!9.13}{[0.292, 0.395]\rlap{$^{\ddagger}$}}\\[0pt]
\addlinespace[1.8pt]
\cellcolor{white}NTv3 650M pre & \unCell{benchCompose!5.03}{[-0.004, 0.021]} & \unCell{benchControl!5.00}{[47.1, 53.9]} & \unCell{benchControl!7.47}{[-0.094, -0.034]} & \unCell{benchControl!10.77}{[0.345, 0.391]}\\[0pt]
\cellcolor{white}NTv3 650M post & \unCell{benchCompose!20.73}{[0.155, 0.220]} & \unCell{benchControl!17.12}{[25.8, 34.5]} & \unCell{benchControl!16.35}{[0.198, 0.377]} & \unCell{benchControl!17.68}{[0.527, 0.610]}\\[0pt]
\cellcolor{white}\unNative{Native outputs} & \unCell{benchCompose!12.20}{[0.138, 0.201]} & \unCell{benchControl!22.65}{[25.5, 33.3]} & \unCell{benchControl!12.98}{[0.162, 0.341]} & \unCell{benchControl!13.84}{[0.430, 0.534]\rlap{$^{\ddagger}$}}\\[0pt]
\addlinespace[2.2pt]
\cellcolor{white}AlphaGenome & \unCell{benchCompose!38.00}{[0.232, 0.304]} & \unCell{benchControl!31.06}{[21.3, 30.3]} & \unCell{benchControl!38.00}{[0.422, 0.612]} & \unCell{benchControl!38.00}{[0.827, 0.839]}\\[0pt]
\cellcolor{white}\unNative{Native outputs} & \unCell{benchCompose!31.61}{[0.219, 0.299]} & \unCell{benchControl!38.00}{[20.5, 28.3]} & \unCell{benchControl!28.69}{[0.288, 0.483]} & \unCell{benchControl!30.15}{[0.701, 0.713]\rlap{$^{\ddagger}$}}\\[0pt]
\cellcolor{white}Enformer & \unCell{benchCompose!15.44}{[0.141, 0.220]} & \unCell{benchControl!12.39}{[27.4, 35.7]} & \unCell{benchControl!21.41}{[0.193, 0.391]} & \unCell{benchControl!24.76}{[0.692, 0.709]\rlap{$^{\ddagger}$}}\\[0pt]
\addlinespace[2.2pt]
\cellcolor{white}$k$-mer ridge & \unCell{benchCompose!5.76}{[0.015, 0.049]} & \unCell{benchControl!6.07}{[41.2, 48.9]} & \unCell{benchControl!6.73}{[-0.185, -0.009]} & \unCell{benchControl!11.91}{[0.368, 0.420]}\\[0pt]
\cellcolor{white}Task-trained ResNet & \unCell{benchCompose!5.00}{[-0.012, 0.011]} & \unCell{benchControl!5.61}{[43.5, 50.6]} & \unCell{benchControl!5.63}{[-0.281, -0.155]} & \unCell{benchControl!9.82}{[0.332, 0.362]}\\[0pt]
\cellcolor{white}SpliceAI & \unEmpty{benchCompose!5.00} & \unEmpty{benchControl!5.00} & \unEmpty{benchControl!5.00} & \unEmpty{benchControl!5.00}\\[0pt]
\addlinespace[3.0pt]
\cellcolor{white}\textcolor{black!65}{Experimental reference} & \unCell{black!3}{[0.381, 0.462]} & \unCell{black!3}{[9.7, 14.0]} & \unCell{black!3}{[0.829, 1.092]} & \unCell{black!3}{[1.598, 1.606]}\\[0pt]
\end{tabular}\end{uncertaintytable}

\begingroup\fontsize{9}{11}\selectfont
\caption{\textbf{Uncertainty of the main results.} Pointwise 95\% confidence
intervals; multi-cell gain shows ranges across the two RNA measurements.
Single values are point estimates. Shading and notation follow
Table~\ref{tab:bend2-reference}; ---: not reported.}
\label{tab:prediction-uncertainty}
\label{tab:design-benchmark}
\endgroup
\end{table}
\FloatBarrier

\end{document}